\documentclass[journal]{IEEEtran}

\usepackage[dvips]{color}
\usepackage{amsmath}
\usepackage{amssymb}
\usepackage{mathrsfs}
\usepackage{graphicx}
\usepackage{upgreek}
\usepackage{bm}
\usepackage{algorithm}
\usepackage{algorithmic}
\usepackage{multirow,makecell}
\usepackage{subfigure}
\usepackage{xcolor}
\usepackage{CJKutf8}
\usepackage{cases}

\usepackage{cite}

\usepackage{xpatch}

\begin{document}					
\title{Robust Transmission Design for  Reconfigurable Intelligent Surface and Movable Antenna Enabled Symbiotic Radio Communications}
					
\author{\IEEEauthorblockN{Bin Lyu,~\IEEEmembership{Senior Member,~IEEE,}~Hao Liu,~Meng Hua,~\IEEEmembership{Member,~IEEE,}~Wenqing Hong,\\~ Shimin Gong,~\IEEEmembership{Member,~IEEE,}~Feng Tian},~and Abbas Jamalipour,~\IEEEmembership{Fellow,~IEEE}

\thanks{This article was presented in part at the IEEE VTC2024-Spring, Singapore, June 24-27, 2024\cite{MYVTC}.

\IEEEcompsocthanksitem Bin Lyu, Hao Liu, Wenqing Hong, and Feng Tian are with Nanjing University of Posts and Telecommunications, Nanjing 210003, China. Meng Hua is with  Imperial College London, London SW7 2AZ, UK. Shimin Gong is with  Shenzhen Campus of Sun Yat-Sen University, Shenzhen 518055, China. Abbas Jamalipour is with University of Sydney, Sydney, NSW 2006, Australia.
}
				
}		

\maketitle
		
\begin{abstract}
This paper explores the application of movable antenna (MA), a cutting-edge technology with the capability of altering antenna positions, in a symbiotic radio (SR)  system enabled by reconfigurable intelligent surface (RIS). The goal is to fully exploit the capabilities of both MA and RIS, constructing a better transmission environment for the co-existing primary and secondary transmission systems. For both parasitic SR (PSR) and commensal SR (CSR) scenarios with the channel uncertainties experienced by all transmission links, we design a robust transmission scheme with the goal of maximizing the primary  rate while ensuring the secondary transmission quality. To address the maximization problem with thorny non-convex characteristics, we propose an alternating optimization framework that utilizes the General S-Procedure, General Sign-Definiteness Principle, successive convex approximation (SCA), and  simulated annealing (SA) improved particle swarm optimization (SA-PSO) algorithms.
 Numerical results validate that the CSR scenario significantly outperforms the PSR scenario in terms of primary  rate, and also show that compared to the fixed-position antenna  scheme, the proposed MA scheme can increase the primary  rate by 1.62 bps/Hz and 2.37 bps/Hz for the PSR and CSR scenarios, respectively.
\end{abstract}
		

\section{Introduction}
\IEEEPARstart{A}{s} released by the International Telecommunication Union, a large-scale, complexly structured, and service-diversified wireless network is expected in the 6G vision \cite{ITU}. To support this impending network, ubiquitous deployment of Internet of Things (IoT) devices is inevitable, resulting in increasingly stricter demands on spectrum and energy resources.
 However, scarce spectrum sources are insufficient to support  spectrum requests from enormous IoT devices, especially for the scenario that each IoT device requests a dedicated spectrum band  \cite{LiangSurvey}. In addition, the mass deployment of IoT devices with  radio frequency (RF) components is generally energy-consuming, which increases the network operating costs and runs counter to  the  goal of carbon neutrality.
These challenges present significant barriers to the evolution of 6G, highlighting the urgent need for innovative solutions that prioritize both spectrum and energy efficiencies.

Symbiotic radio (SR) has recently emerged to address the aforementioned challenges by  establishing a mutually beneficial relationship between primary and secondary transmissions \cite{LongIoT,Zeng,QQZhang,ZengMassive}. Specifically, IoT devices constituting secondary transmission, also called secondary devices, are capable of sending their data by riding along with primary signals over the spectrum band reserved for the primary transmission. 
 Compared to these devices based on RF components, the mass deployment of secondary devices  can play a significant role in energy conservation. The concept of SR communication was systematically described in \cite{LongIoT}, in which the classical parasitic SR (PSR) and commensal SR (CSR) scenarios were proposed. {These two scenarios are mainly defined based on the symbol period ratio between primary and secondary transmissions \cite{LongIoT}, which introduce different characteristics of SR communications. In particular, the symbol periods of primary and secondary signals for the PSR scenario are the same. While, the secondary signal for the CSR scenario has a much larger period than the primary signal.}
The authors in \cite{Zeng} characterized the tradeoff between primary and secondary  rates for  the PSR scenario, and the condition of attaining mutualistic symbiosis for the CSR scenario was revealed in \cite{QQZhang}.  The authors in \cite{ZengMassive}  demonstrated the positive effect of the number of secondary devices.  However, secondary devices in \cite{LongIoT,Zeng,QQZhang,ZengMassive} are generally backscatter devices with limited reflecting capabilities, which cannot meet the requirement of high transmission rates.

Reconfigurable intelligent surface (RIS) has been widely used in performance improvement of wireless communications  \cite{risone,HuaRIS,RISSurvey}. RIS can be seen as an evolutionary version of backscatter devices, which has massive elements to adjust amplitudes and phase shifts of incident signals and achieve the corresponding signal reflections. Inspired by this, it makes sense to incorporate RIS into SR systems \cite{MengHua,HuaUAV,Zhou,BPSK,LYUBIN,YongjunRSMA}. In \cite{MengHua}, an RIS enabled SR system was modeled, based on which the system bit error rate   is minimized for both PSR and CSR scenarios. In \cite{HuaUAV}, an unmanned aerial vehicle (UAV) served as the primary transmitter (PT) in RIS enabled SR systems and leveraged its mobility to rebuild favorable transmission conditions. 
 To support the primary transmission and establish the secondary transmission concurrently, a novel RIS partitioning technique was created in \cite{Zhou}.  In \cite{BPSK}, multiple antennas were exploited at the transceivers for performance enhancement by utilizing the antenna diversity. 
Note that the investigations of \cite{MengHua,HuaUAV,Zhou,BPSK} were based on the perfect acquisition of channel state information (CSI). However, estimation mistakes  throughout the acquisition process may result in an inevitability of CSI uncertainties \cite{Ng}. In light of this, robust designs for SR systems by taking into account channel uncertainties are urgent. \cite{LYUBIN} adopted the separated CSI model and proposed a robust beamforming scheme for both PSR and CSR scenarios to suppress eavesdropping from the eavesdroppers.
However, the separated CSIs associated with secondary devices (i.e., backscatter devices or RISs)  are challenging to obtain independently since secondary devices are passive for neither actively sending nor receiving pilot signals \cite{Pan}. Thus, the estimation of cascaded channels has been considered in more studies \cite{PanTwo}, resulting in a cascaded CSI error model \cite{cascaded} for the associated uncertainty. To improve the robustness of RIS enabled SR systems, the beamforming was designed for the CSR scenario  by considering both direct and cascaded CSI uncertainties  in \cite{YongjunRSMA}. 
{Despite the efforts of utilizing RIS in SR systems have improved  system performance \cite{MengHua,HuaUAV,Zhou,BPSK,LYUBIN,YongjunRSMA}, there still exist gaps to explore the full potential of system performance. For example, due to the fixed-position antennas (FPAs) applied at transceivers, the full utilization of spatial degrees of freedom (DoFs) related with the direct and cascaded links cannot be achieved. In particular, for the direct link, its transmission condition  cannot be improved by the RIS, which results in a gap for achieving satisfactory primary transmission performance for the PSR and CSR scenarios, especially for the PSR scenario.

Recently,  an innovative technology, movable antenna (MA) \cite{MAMag,MATWC}, also known as fluid antenna \cite{FAS},  has offered a viable solution to fully leverage the DoFs for system performance enhancement \cite{Lipeng,Weidong,QQWu,MAthree,PSOMA,Pengcheng,MA-TMC,TwoScale,MoveDelay}.} The MA technology, in contrast to the FPAs, use flexible cables as the connections between antennas and matching RF chains. This allows antennas to dynamically change their positions within a specific range in order to create better communication environments \cite{MAMag}. In \cite{MATWC} and \cite{Lipeng}, the concept of MA was first systematically introduced and a field-response model was developed to characterize the channels associated with MAs at transceivers. After that, \cite{Weidong} utilized the MAs to boost the received power and designed a graph-based approach to find the optimal positions of MAs. {In \cite{QQWu}, the MA was applied in a multiple-input single-output interference network to suppress the inter-cell interference, while ensuring the desired signal strength.}  In \cite{MAthree}, the combination of MAs and non-orthogonal multiple access (NOMA) was investigated to improve the system sum rate and reduce the outage probability. In \cite{MA-TMC}, an MA empowered approach was designed to achieve secure transmissions between the transceivers, resisting the eavesdropping from  multiple eavesdroppers. {In \cite{MoveDelay}, the time delay caused by MA movement was first considered as a factor for performance optimization. Unlike  \cite{Lipeng,Weidong,QQWu,MAthree,MA-TMC,MoveDelay}, \cite{PSOMA} and \cite{TwoScale} focused on multi-user uplink communication systems empowered by MAs. In particular, \cite{PSOMA} exploited the particle swarm optimization (PSO) algorithm to attain the optimized MA positions. In \cite{TwoScale}, a novel two-scale design was proposed to balance the system performance and practical implementation constraints.} \cite{Pengcheng} utilized the MAs to boost downlink energy transfer as well as uplink task offloading in an effort to increase computational capabilities while utilizing fewer resources. 

Driven by this superior performance brought by the MA technology, integrating MAs into SR systems appears to be a potential move \cite{ZHOUMASR,guanjiayu}. In \cite{ZHOUMASR}, a one-dimensional MA array was leveraged to  construct the mutualism of primary and secondary transmission systems for the CSR scenario. In \cite{guanjiayu}, the potential of MAs was exploited in SR  systems with eavesdroppers with the goal of   amplifying the signal strength at preferred directions and preventing the undesired eavesdropping, creating a new avenue for secure transmission. {However, the considered SR  systems with MAs in \cite{ZHOUMASR} and \cite{guanjiayu} undergo
an obvious shortcoming that  the secondary device is the backscatter device with one single antenna.  
 In the meanwhile, the results attained in \cite{ZHOUMASR} and \cite{guanjiayu} were also based on the perfect CSI assumption, which is impractical and leads to the mismatch of transmission designs. 

 To address these drawbacks encountered by \cite{MengHua,HuaUAV,Zhou,BPSK,LYUBIN,YongjunRSMA} and \cite{ZHOUMASR,guanjiayu}, the application of MA in RIS enabled SR  systems 
is an efficient solution, which has 
not been studied yet, leading to a gap in constructing better channel conditions for both PSR and CSR scenarios. In particular, how to apply the MA to enhance the direct link condition to deliver primary signals and how to jointly apply the MA and RIS to balance the performance of primary and secondary transmissions are unknown.
Furthermore, designing the robust transmission scheme against the CSI uncertainties for the RIS enabled SR  system with MAs is also unresolved. 
To  bridge the aforementioned gaps, in an SR  system, we explore the application of multiple MAs at the PT to reconfigure its channel conditions  by manipulating the MA positions within a three-dimensional region, and leverage the RIS as the secondary device to support the secondary transmission by adjusting its phase shifts.
This configuration allows us to take advantage of additional DoFs provided by the MAs and the channel reconfiguration characteristics provided by the RIS for  performance enhancement.} Then, we propose a robust transmission scheme for the PSR and CSR scenarios to combat the CSI  uncertainties  experienced by both direct and cascaded links, respectively. Compared to \cite{MengHua,HuaUAV,Zhou,BPSK} in which the PT is with the FPAs, having multiple MAs in our model results in a challenging optimization of transmit beamforming and MA positions. Unlike \cite{LYUBIN}, we consider the full channel uncertainties for a more robust design. Different from \cite{YongjunRSMA}, the extended investigation of the PSR scenario with MAs is more challenging, and the proposed scheme in \cite{YongjunRSMA} is also inappropriate for the CSR scenario in our model. In contrast to \cite{MAMag,MATWC,Lipeng,Weidong,QQWu,MAthree,MA-TMC,MoveDelay,PSOMA,TwoScale}, 
the quality of service (QoS) for primary and secondary transmissions should be balanced by our robust beamforming design. Moreover, the robust design and the comparison between PSR and CSR scenarios are also missing in the previous works, i.e.,~\cite{ZHOUMASR} and \cite{guanjiayu}. We summarize the main contributions of this paper as follows:

\begin{itemize}
	\item{This is the first work with the utilization of MA in RIS enabled SR systems, aiming at leveraging the potential of MA and RIS technologies.  For the PSR scenario, the MA can be utilized to create a more effective  environment for the transmission of primary signals, and the MA as well as the RIS are jointly adopted to improve the efficiency of secondary transmission. For the CSR scenario, both primary and  secondary transmissions can be boosted by the joint application of the MA and RIS, and the corresponding transmission requirements can be balanced by an appropriate adjustment of the MA positions and RIS phase shifts. 
	 }
	\item{Considering the presence of direct and cascaded CSI uncertainties,  we look into the robust  transmission scheme  for the PSR and CSR scenarios. Specifically, for each scenario, we maximize the minimum primary rate  with the constraint on secondary transmission performance to ensure the worst-case transmission robustness. 
We provide an alternating optimization (AO) framework for obtaining a near-optimal solution to the original problem.
For the transmit beamforming design, the utilization of the successive convex approximation (SCA) method in conjunction with the General S-Procedure and the General Sign-Definiteness Principle is considered. 
To optimize the passive beamforming, the intractable discrete phase shift constraint is approximated into tractable forms by introducing binary variables and utilizing the SCA method. 
After that, the simulated annealing (SA) improved particle swarm optimization (SA-PSO) algorithm is applied to optimize the MA positions, ensuring a satisfactory solution accuracy.}


	\item{We conduct numerical results to assess the performance improvement realized by  our proposed scheme. Specifically, compared to the scheme with deploying FPAs at the PT, our proposed scheme can attain 12.1\% and 6.6\% performance gains in terms of primary rate for the PSR and CSR scenarios, respectively. Moreover, we demonstrate the performance superiority of the CSR scenario over the PSR scenario.}
	\end{itemize}


The rest of the paper is organized as follows. We present the system model in Section \ref{moxing}. Section \ref{sanpsr} and Section \ref{sicsr} describe the investigations of primary rate maximization problems for the PSR and CSR scenarios, respectively. Section \ref{results} provides numerical simulation results for performance verification. The conclusion of the paper is outlined in Section \ref{Conc}.

$\textbf{Notations:}$ Vectors and matrices are represented as boldface lowercase and boldface uppercase letters, respectively. $||\bm{x} ||_2$ is the L$_2$-norm of $\bm{x}$.
 \text{diag}${(\bm{x})}$ constructs a  matrix with the diagonal vector $\bm{x}$.
  $\bm{X}^T$, $\bm{X}^*$, and $\bm{X}^H$ represent the transpose, conjugate, and Hermitian operations of $\bm{X}$, respectively. The Frobenius norm of $\bm{X}$ is symbolized by $||\bm{X}||_F$. $\bm{X} \succeq \bm{Y}$ represents that $\bm{X} - \bm{Y}$ is a positive semidefinite matrix. $\bm{X} \otimes \bm{Y}$ denotes the kronecker product between them.
  $\text{vec}(\bm{X})$ is the vectorization operation of $\bm{X}$.

\section{System Model}
\label{moxing}
\begin{figure}[t]
  \centering
  \includegraphics[width=0.38\textwidth]{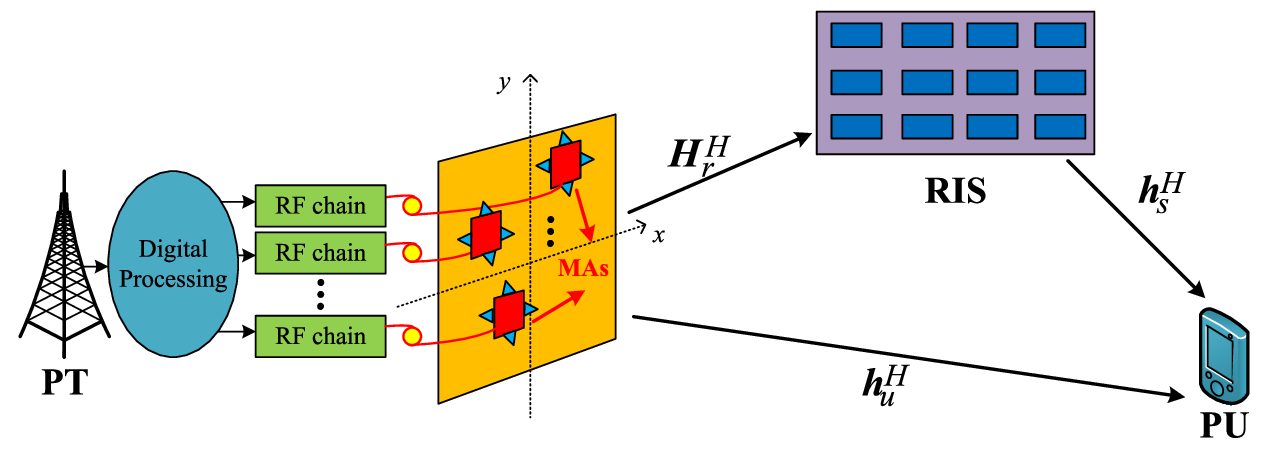}
  \caption{An RIS enabled SR communication system with MAs.}
  \label{fig:figure_label}
\end{figure}

An RIS enabled SR communication system with MAs is shown 
in Fig.~\ref{fig:figure_label}, which is composed of a PT having $K$ MAs, an RIS having $M$ reflecting elements with fixed positions and a PU with single FPA.\footnote{This system model can be easily expanded to the multi-PU scenario and the detailed discussions can be found in Section \ref{sub:multi_pu_scenario}.} The RIS is leveraged to support two functionalities for SR communications. First, the RIS can provide additional links to enhance the transmission quality from the PT to the PU\footnote{For the case that the direct link  is unavailable, the ambiguity problem of jointly decoding primary and secondary signals may happen. To avoid this dilemma, the RIS partitioning scheme proposed in \cite{Zhou} can be used to partition the original surface  into two sub-surfaces, where one sub-surface is used to establish a relaying link for exclusively transmitting the primary signals, and the other sub-surface is used to achieve the secondary transmission.}. Second, the RIS delivers its detected environmental information, e.g., atmospheric carbon dioxide levels and humidity, by reusing primary signals emitted by the PT. The movable capability of MAs is provided by using flexible cables to connect their corresponding RF chains, based on which each MA can move itself to a favorable position \cite{MAMag,MATWC}. Denote by $\bm{p}_{k} =[x_{k}, y_{k},z_{k}]^T \in \mathcal{C}$ the position of the $k$-th MA, where $\mathcal{C}$ is  constrained by a  3D region with size $[x_\text{min},x_\text{max}]^K \times [y_\text{min}, y_\text{max}]^K \times [z_\text{min}, z_\text{max}]^K$, $k \in \mathcal{K}$, and $\mathcal{K}=\{1,2,\ldots,K\}$ represents the set of all MAs. Denote by  $\bm{p}_{r,m} =[x_{r,m}, y_{r,m}, z_{r,m}]^T$ and $\bm{p}_{u} =[x_u, y_u, z_u]^T$ the reference positions of the RIS and the PU, respectively,  where $m \in \mathcal{M}$, and $\mathcal{M} = \{1,2,\ldots,M\}$.

\subsection{Channel Model}
 Denote by $\bm{H}_{r}^H \in \mathbb{C}^{\ M\times K} $, $\bm{h}_{u}^H \in \mathbb{C}^{\ 1\times K} $, and $\bm{h}_{s}^H \in \mathbb{C}^{\ 1\times M} $ the channels from the PT to the RIS, the PT to the PU, and the RIS to the PU,  respectively. {To describe  the multi-path propagation environment associated with the MAs, we adopt the widely-used far-field response channel model developed by \cite{MATWC} and \cite{Lipeng}, which consists of parameters such as amplitude, phase, angle of arrival (AoA) and angle of departure (AoD) of each channel path. The far-field characteristic of this channel model holds  due to the fact the  transmission distance between the PT and the RIS/PU is much greater than the size of the transmit region at the PT, based on which the amplitudes, AoAs and AoDs keep unchanged when the MA positions are updated, while the phases are variables related to the MA positions. }

 Define the set of the RIS and the PU as $\kappa =\{r,u\}$, and denote by $L_{\kappa}^{t}$ and $L_{\kappa}^{r}$ the amounts of the transmit and receive paths between the PT and the RIS/PU, respectively.  The transmit azimuth and elevation angles of departure (AoDs), as well as  the receive azimuth and elevation angles of arrival (AoAs) between the PT and the RIS/PU are $\phi_{\kappa,\breve{\iota}}^t \in [0,\pi]$, $\theta_{\kappa,\breve{\iota}}^t \in [0,\pi]$, $1\le \breve{\iota} \le L_{\kappa}^t$ and $\phi_{\kappa,\iota}^r \in [0,\pi]$, $\theta_{\kappa,\iota}^r \in [0,\pi]$, $1\le \iota \le L_{\kappa}^r$, respectively. 

The signal propagation difference for the RIS/PU between the $\breve{\iota}$-th transmit path of the $k$-th MA and the origin of the transmit region is expressed as  $\rho_{\kappa,\breve{\iota}}^t(\bm{p}_k) = x_{k} \cos\theta_{\kappa,\breve{\iota}}^t \cos\phi_{\kappa,\breve{\iota}}^t + y_{k} \cos \theta_{\kappa,\breve{\iota}}^t \sin \phi_{\kappa,\breve{\iota}}^t + z_{k} \sin \theta_{\kappa,\breve{\iota}}^t, ~\kappa = \{r,u\}.$
Accordingly, the transmit-field response matrix at the PT is expressed as $\bm{G}_\kappa= [ \bm{g}_\kappa(\bm{p}_1), \ldots, \bm{g}_\kappa(\bm{p}_K) ] \in \mathbb{C}^{ L_{\kappa}^t \times K }$, where $\bm{g}_\kappa (\bm{p}_k) = [e^{j \frac{2\pi}{\lambda}\rho_{\kappa,1}^t}, \ldots, e^{j \frac{2\pi}{\lambda}\rho_{\kappa,L_{\kappa}^t}^t}  ]^T \in \mathbb{C}^{L_{\kappa}^t \times 1}$, and $\lambda$ is the carrier wavelength.
The signal propagation difference between the $\iota$-th receive path of the $m$-th RIS element and the origin of the reference point is expressed as
$\rho_{r,\iota}^r(\bm{p}_{r,m}) = x_{r,m} \cos\theta_{r,\iota}^r \cos\phi_{r,\iota}^r + y_{r,m} \cos \theta_{r,\iota}^r \sin \phi_{r,\iota}^r + z_{r,m} \sin \theta_{r,\iota}^r$. Then, the receive-field response matrix at the RIS is modeled as $\bm{F}_{r} = [ \bm{f}_{r}(\bm{p}_{r,1}), \ldots, \bm{f}_{r}(\bm{p}_{r,M}) ] \in \mathbb{C}^{L_{r}^r \times M}$, where $\bm{f}_{r}(\bm{p}_{r,m}) = [e^{j \frac{2\pi}{\lambda}\rho_{r,1}^r}, \ldots, e^{j \frac{2\pi}{\lambda}\rho_{r,L_{r}^r}^r}  ]^T \in \mathbb{C}^{L_{r}^r \times 1}$ represents the receive-field response vector associated with the $m$-th element of the RIS. Similarly, the  receive-field response vector at the PU associated with the PT is $\bm{f}_{u} = [e^{j \frac{2\pi}{\lambda}\rho_{u,1}^r}, \ldots, e^{j \frac{2\pi}{\lambda}\rho_{u,L_{u}^r}^r}  ]^T \in \mathbb{C}^{L_{u}^r \times 1}$, where $\rho_{u,\iota}^r(\bm{p}_{u}) = x_{u} \cos\theta_{u,\iota}^r \cos\phi_{u,\iota}^r + y_{u} \cos \theta_{u,\iota}^r \sin \phi_{u,\iota}^r + z_{u} \sin \theta_{u,\iota}^r$. Thus, the channels associated with the PT are expressed as $\bm{H}_{r}^H = \bm{F}_{r}^H \bm{\Sigma}_{r} \bm{G}_r$ and $\bm{h}_{u}^H = \bm{f}_{u}^H \bm{\Sigma}_{u} \bm{G}_{u}$, where $\bm{\Sigma}_{\kappa} \in \mathbb{C}^{L_{\kappa}^r \times L_{\kappa}^t}$ is the path-response matrices from the PT to the RIS/PU.

Let $\bm{h}_s^H$ represent the channel from the RIS to the PU and be modeled as $\bm{h}_s^H = \bm{f}_{s}^H \bm{\Sigma}_{s} \bm{G}_{s}$, where $\bm{G}_{s}$ is the transmit-field response matrix at the RIS, $\bm{\Sigma}_{s}$ is the associated path-response matrix, and $\bm{f}_{s}$ is the associated receive-field response vector.
Define the cascaded channel from the PT to the PU via the RIS as $\bm{H}_{bs}$, which  is expressed as
$\bm{H}_{bs} = \text{diag}(\bm{h}_{s}^H) \bm{H}_{r}^H$.
Due to the characteristic of  MAs, $\bm{H}_{bs}$ and $\bm{h}_{u}^H$ are variables and better CSI can be guaranteed by  updating the MA positions.  However, $\bm{h}_s^H$ is invariant in a considered transmission block due to the utilization of FPAs at transceivers.

\subsection{Full Channel Uncertainty}
It is known that perfect CSI in RIS enabled SR systems is generally challenging to acquire and estimation errors are unavoidable \cite{cascaded}. Thus, to investigate  how the estimation errors affect  system performance and present robust transmission schemes, we consider that both the cascaded and direct channels are imperfect. In light of this, the bounded CSI error model \cite{cascaded} is utilized as follows\footnote{We can also consider the statistical CSI error model to measure the full channel uncertainty, based on which the system performance optimization problem can be formulated as a transmit power minimization problem under the outage probability constraints on primary and secondary transmissions. To solve this optimization problem, we can modify the proposed solutions shown in Section \ref{sanpsr} and Section \ref{sicsr} with utilizing the Bernstein-Type Inequality \cite{Bernstein-Type} and semi-definite relaxation. However, due to the limited space, the detailed study will be shown in the future work.}:
\begin{align}
\label{imperfectchannel}
&\bm{H}_{bs} = \widehat{\bm{H}}_{bs} + \Delta \bm{H}_{bs},~~~~ ||\Delta \bm{H}_{bs}||_F \le \xi_{bs},\\
\label{imperfectdirectchannel}
&\bm{h}_{u} = \widehat{\bm{h}}_{u} + \Delta \bm{h}_{u},~~~~~~~~~ ||\Delta \bm{h}_{u}||_2 \le \xi_{u},
\end{align}
where $\widehat{\bm{H}}_{bs}$ and $\widehat{\bm{h}}_{u}$ are the estimated CSI values and known by the PT. $\Delta \bm{H}_{bs}$ and $\Delta \bm{h}_{u}$ represent the uncertainties.  {$\xi_{bs}$ and $\xi_{u}$ are the radii of the uncertainty regions and can be acquired by taking into account the estimation error, quantization error \cite{PanTwo} and outdated feedback \cite{Uncertainty}.}

\subsection{Transmission Model}
Denote by $s(l)$ and $\bm{w} \in \mathcal{C}^{K\times 1}$ the primary signal and the transmit beamforming vector at the PT, respectively, where $s(l)\sim \mathcal{CN}(0,1)$ and $||\bm{w}||^2\le P_\text{max}$, and $P_\text{max}$ is the upper limit of the transmit power. {We consider both PSR and CSR scenarios, which are suitable for different applications and cannot be replaced by each other. Specifically, the PSR scenario is suitable for the applications that require high-quality secondary transmission but low primary rates, while the CSR scenario is applicable to the opposite applications.} For both PSR and CSR scenarios, the expressions of the incident signal at the RIS from the PT are the same and  denoted by $\bm{H}^H_r \bm{w}s(l)$.  Then, the transmission models can be discussed as follows: 

 \subsubsection{PSR scenario}
 {The secondary signal for the PSR scenario, denoted by $c(l)$, shares the same symbol duration with $s(l)$.} {Considering the hardware limitation encountered by the RIS, we use the binary phase shift keying  scheme as it is easy for  practical implementation \cite{Zhou,BPSK}, which is with equal probability of sending symbol `1' and symbol `-1', yielding $c(l) \in \{-1,1\}$ \cite{BPSK,LYUBIN}.}
 The $l$-th  signal received by the PU is given by
\begin{align}
\label{ReceivedSignal}
y_{psr,r}(l) = \bm{h}_u^H \bm{w}s(l) + \bm{h}_s^H \bm\Theta \bm{H}_r^H \bm{w} s(l) c(l)+z(l),
\end{align}
where $\bm\Theta = \text{diag}(e^{j\theta_1},\ldots, e^{j\theta_M})$ is the RIS phase-shift matrix,  $\theta_m \in \left\{0,\frac{2\pi}{\overline\kappa}, \ldots, \frac{2\pi(\overline\kappa - 1)}{\overline\kappa}\right\}$ represents the discrete  phase-shift of the $m$-th element, $\overline\kappa$
is the number of quantization level, and $z(l) \sim \mathcal{CN}(0,\sigma^2)$ denotes the additive Gaussian white noise. {Exploiting the characteristic of the PSR,  we first decode $s(l)$ by treating the secondary signal part as interference, owing to the coupling of $s(l)$ and $c(l)$ \cite{LongIoT}.} When decoding $s(l)$, the signal-to-interference-plus-noise ratio (SINR) is expressed as 
\begin{align}
\label{PSRSNR}
\gamma_{psr,s} = \frac{|\bm{h}_u^H \bm{w}|^2 } {|\bm{h}_s^H  \bm\Theta \bm{H}_r^H \bm{w}|^2 + \sigma^2 } = \frac{|\bm{h}_u^H \bm{w}|^2 }{|\bm{\psi}^H \bm{H}_{bs} \bm{w}|^2 + \sigma^2 }, 
\end{align}
where $\bm{\psi}=  [\psi_1, \ldots,\psi_M]^T $ with $\psi_m = e^{j\theta_m}$ for $m\in \mathcal{M}$, which is also known as the passive beamforming vector. To evaluate the worst-case interference caused by the secondary transmission, we assume that the reflected signal $s(l)$$c(l)$ satisfies the circularly symmetric complex Gaussian distribution \cite{LongIoT}. Thus, the lower bound of the primary  rate is expressed as
\begin{align}
  &R_{psr} = \log_2(1+\gamma_{psr,s}) = \log_2(1+\frac{|\bm{h}_u^H \bm{w}|^2 }{|\bm{\psi}^H  \bm{H}_{bs} \bm{w}|^2 + \sigma^2 }). 
\end{align}
After decoding $s(l)$, we can derive the signal-to-noise ratio (SNR) for decoding $c(l)$ as $\gamma_{psr,c} = \frac{|\bm{\psi}^H \bm{H}_{bs} \bm{w}|^2}{\sigma^2}$ by the successive interference cancellation.

 \subsubsection{CSR scenario} 
{The secondary signal for the CSR scenario, denoted by $c$, can scan $L$ primary signals (also $L$ secondary signals for the PSR scenario), i.e., $ [ s(1), \dots, s(L)]$, where $L \gg 1$.} {We also assume  $c \in \{-1,1\}$ with equal probability \cite{BPSK,LYUBIN}.} 
 In this scenario, the $l$-th signal received by the PU is given by
\begin{align}
\label{ReceivedSignalCSR}
y_{csr,r}(l) = \bm{h}_u^H \bm{w}s(l) + \bm{h}_s^H  \bm\Theta \bm{H}_r^H \bm{w} s(l) c +z(l).
\end{align}
{Different from the PSR scenario, $\bm{h}_s^H  \bm\Theta \bm{H}_r^H \bm{w} s(l) c$ in $\eqref{ReceivedSignalCSR}$ is no longer the interference term when decoding $s(l)$ as it can be viewed as a multipath element for delivering $s(l)$~\cite{LongIoT}.} Accordingly, the SNR for decoding $s(l)$ is given by 
\begin{align}
\label{CSRSNR}
\gamma_{csr,s} = {|\bm{h}_u^H \bm{w} + \bm{h}_s^H  \bm\Theta \bm{H}_r^H \bm{w} c|^2 }/{\sigma^2 },
\end{align}
{and the primary rate can be derived by taking the expectation of $c$ and shown as follows:
\begin{align}
 R_{csr} &= \nonumber\mathbb{E}_c [ \log_2(1+\gamma_{csr,s})  ] \\\nonumber&={\frac{1}{2}\log_2(1+{|\bm{h}_u^H \bm{w} + \bm{h}_s^H  \bm\Theta \bm{H}_r^H \bm{w} |^2 }/{\sigma^2 })} \\ &+ {\frac{1}{2}\log_2(1+{|\bm{h}_u^H \bm{w} - \bm{h}_s^H  \bm\Theta \bm{H}_r^H \bm{w} |^2 }/{\sigma^2 })},
\end{align}
where the first term corresponds to the case that the RIS sends symbol `1' with probability $\frac{1}{2}$, and the second term corresponds to  the case that the RIS sends symbol `-1' with probability $\frac{1}{2}$.}

By subtracting $\bm{h}_u^H \bm{w}s(l) $ from $y_{csr,r}(l)$ and leveraging the maximum ratio combing~\cite{BPSK}, the SNR for decoding $c$ for the CSR scenario is formulated as $\gamma_{csr,c} \approx  \frac{L |\bm{\psi}^H \bm{H}_{bs} \bm{w}|^2}{\sigma^2}$.


\section{Robust Transmission Design for PSR}
\label{sanpsr}
Our goal is to maximize the worst-case of the primary transmission performance for the PSR scenario under the QoS constraint on the  secondary signal decoding. This can be  achieved by investigating the robust design of transmit beamforming $\bm{w}$, the passive beamforming $\bm{\psi}$, and the MA positions $\bm{p}= [\bm{p}_1^T,\ldots, \bm{p}_K^T]^T$. Thus, we can formulate the maximization problem as
\begin{equation}\tag{$\textbf{P1}$}
		\begin{aligned}
			\max_{\bm{w}, \bm{\psi}, \bm{p}} ~~& [\begin{array}{c}
   \mathop{\min}\limits_{\Delta{\bm{H}_{bs}}, \Delta{\bm{h}_u}}
     \{  \log_2(1+\gamma_{psr,s})  \}
 \end{array} ] \\
			\text{s.t.}~~ & \text{C1:~} ||\bm{w} ||_2^2 \le P_\text{max}, \\
			& \text{C2:~} \gamma_{psr,c} \ge \gamma_\text{pmin}, ~||\Delta \bm{H}_{bs}||_F \le \xi_{bs},\\
                & \text{C3:~} \psi_m = e^{j \theta_m}, ~ m \in \mathcal{M},\\
			& \text{C4:~} x_\text{min}\le x_k \le x_\text{max},
			  ~y_\text{min}\le y_k \le y_\text{max},\\
     &~~~~~z_\text{min}\le z_k \le z_\text{max}, ~ k \in \mathcal{K},\\
			& \text{C5:~} \left\|\bm{p}_k - \bm{p}_o \right \|_2 \ge d_\text{min},~k, o \in \mathcal{K},~k\neq o, \\
		\end{aligned}
	\end{equation}
where $\gamma_\text{pmin}$ represents the minimum SNR required when decoding $c(l)$ for the PSR scenario,   and $d_\text{min}$ represents the minimum distance between  two adjacent MAs. {Note that $\mathop{\min}\limits_{\Delta{\bm{H}_{bs}}, \Delta{\bm{h}_u}} \{  \log_2(1+\gamma_{psr,s})  \}$ represents the minimum value of all  possible primary rates for the PSR scenario within the CSI uncertainty region  defined in \eqref{imperfectchannel} and \eqref{imperfectdirectchannel}, which corresponds to the worst-case of the primary transmission performance.  
To ensure the worst-case robustness, problem \textbf{P1} is formulated as the maximization of $\mathop{\min}\limits_{\Delta{\bm{H}_{bs}}, \Delta{\bm{h}_u}} \{  \log_2(1+\gamma_{psr,s})  \}$. This max-min form has been widely used in the literature for robust design under the bounded CSI error model \cite{Ng}.} C1 restricts the transmit power to not exceed $P_\text{max}$, C2 is the secondary transmission QoS constraint under the bounded CSI error model, C3 defines the feasible set of phase shifts (i.e., passive beamforming), C4 limits the movable range of MAs, and C5 specifies that the distance restriction between two adjacent MAs to avoid the coupling effect.

 \textbf{P1} exhibits a high level of coupling among the variables $\bm{w}$, $\bm\psi$ and $\bm{p}$, rendering it a non-convex optimization that complicates the finding of an optimal solution.
To handle  this, we present an alternating optimization (AO) framework that updates $\bm{w}$, $\bm\psi$ and $\bm{p}$ independently and sequentially. In the steps involving the optimization of $\bm{w}$ and $\bm{\psi}$, we employ the successive convex approximation (SCA) method to handle the non-convexity, and incorporate the General S-Procedure  \cite{ssp} and the General Sign-Definiteness Principle  \cite{lemmatwo}  to effectively address the intractable forms caused by the channel uncertainties. Furthermore, in order to optimize $\bm{p}$ and determine the MA positions for achieving the maximum primary rate, we utilize the SA-PSO algorithm \cite{HPSO}.

        \subsection{Transmit Beamforming Optimization}
	\label{OpTB}
We first maximize $\bm{w}$ by fixing $\bm {\psi}$ and $\bm{p}$. The corresponding sub-problem reduced from \textbf{P1} is expressed as
\begin{equation}\tag{$\textbf{P2}$}
		\begin{aligned}
			\max_{\bm{w}}  &  \left[\begin{array}{c}
   \mathop{\min}\limits_{\Delta{\bm{H}_{bs}}, \Delta{\bm{h}_u}}
     \{  \log_2(1+\gamma_{psr,s})  \}
 \end{array} \right] ~~ 
			\text{s.t.} & \text{C1},~\text{C2}.
		\end{aligned}
	\end{equation}
The objective function is a max-min operation involving the channel uncertainties, which is intractable. Hence, we first derive a lower bound of the objective function as follows \cite{logzeng}:
   $\mathop{\min}\limits_{\Delta{\bm{H}_{bs}}, \Delta{\bm{h}_u}}
        \{  \log_2 (1+\frac{|\bm{h}_u^H \bm{w}|^2 }{|\bm{\psi}^H  \bm{H}_{bs} \bm{w}|^2 + \sigma^2 } )  \}  \geq   \log_2\left(1+ \frac{\mathop{\min}\limits_{\Delta{\bm{h}_u}} \left\{ |\bm{h}_u^H \bm{w}|^2\right\}}{\mathop{\max}\limits_{\Delta{\bm{H}_{bs}}}  \left\{|\bm{\psi}^H  \bm{H}_{bs} \bm{w}|^2 + \sigma^2 \right\}} \right)$.
This allows us to reformulate problem \textbf{P2} as follows
\begin{equation}\tag{$\textbf{P2.1}$}
		\begin{aligned}
			\max_{\bm{w}, \varepsilon_u, \varepsilon_H } ~~&   \log_2(1+\frac{\varepsilon_u}{\varepsilon_H  + \sigma^2})
  \\
			\text{s.t.}~~ & \text{C1},~\text{C2}, \\
                &\text{C6}:~ \varepsilon_u + \mathop{\min}\limits_{ \Delta{\bm{h}_u}} \left\{- {|\bm{h}_u^H \bm{w}|^2} \right\} \leq 0, \\
                &\text{C7}:~ \mathop{\max}\limits_{\Delta{\bm{H}_{bs}} }  \left\{|\bm{\psi}^H  \bm{H}_{bs} \bm{w}|^2 \right\} \leq \varepsilon_H,
		\end{aligned}
	\end{equation}
where $\varepsilon_u$ and $\varepsilon_H$ are two slack variables. Next, to tackle the non-convexity of the objective function in \textbf{P2.1}, we first equivalently reformulate it as a difference of convex form \cite{Ng}:
		\begin{align}
			\nonumber &\max_{\bm{w}, \varepsilon_u, \varepsilon_H } ~~   \log_2(1+\frac{\varepsilon_u}{\varepsilon_H  + \sigma^2})
			\\\Longleftrightarrow &\max_{\bm{w}, \varepsilon_u, \varepsilon_H } ~  {Z}_{fun}(\varepsilon_u, \varepsilon_H) - {B}_{fun}( \varepsilon_H),
		\end{align}
where ${Z}_{fun}(\varepsilon_u, \varepsilon_H) = \log_2(\varepsilon_u + \varepsilon_H + \sigma^2)$	and ${B}_{fun}( \varepsilon_H) = \log_2(\varepsilon_H +\sigma^2)$. However, this form is still non-convex. We thus utilize the SCA method to derive the lower bound of  $B_{fun}(\varepsilon_H)$:
\begin{align}
   \nonumber {B}_{fun}( \varepsilon_H) &\leq  \frac{\varepsilon_H - \varepsilon_H^{(n)}}{(\ln2)(\sigma^2 + \varepsilon_H^{(n)})} + \log_2(\varepsilon_H^{(n)} +\sigma^2),
		\end{align}
where $n$ is the iteration index of the SCA method.
Then, the objective function of \textbf{P2.1} is updated as follows
\begin{align}
\log_2(\varepsilon_u + \varepsilon_H + \sigma^2) - \frac{\varepsilon_H - \varepsilon_H^{(n)}}{(\ln2)(\sigma^2 + \varepsilon_H^{(n)})} - \log_2(\varepsilon_H^{(n)} +\sigma^2).\nonumber
	\end{align}

The constraints C2, C6, and C7 also suffer from the non-convexity. We first substitute $\bm{H}_{bs} = \widehat{\bm{H}}_{bs} + \Delta \bm{H}_{bs}$ into  $\text{C2}$ and adopt the SCA method to approximate  $|\bm{\psi}^H {(\widehat{\bm{H}}_{bs} + \Delta \bm{H}_{bs})}   \bm{w}|^2$ by its lower bound at  ${(\bm{w}^{(n)}, \bm{\psi})}$~\cite{cascaded}:
\begin{align}
\text{vec}^T{(\Delta \bm{H}_{bs})} \bm{A} \text{vec}{(\Delta \bm{H}_{bs} ^*)} + 2\text{Re}\left\{ \bm{a}^T   \text{vec}{(\Delta \bm{H}_{bs} ^*)} \right\} + a, \nonumber
\end{align}
where $\bm{w}^{(n)}$ is the solution obtained at the $n$-th iteration  of using the SCA method to solve \textbf{P2.1},  $\bm{A} = \bm{w}\bm{w}^{(n),H} \otimes \bm{\psi}^*\bm{\psi}^{T} + \bm{w}^{(n)}\bm{w}^{H} \otimes \bm{\psi}^{*}\bm{\psi}^{T} - \left( \bm{w}^{(n)}\bm{w}^{(n),H} \otimes \bm{\psi}^{*}\bm{\psi}^{T} \right)$, $a=$
$2\text{Re}\left\{ \bm{\psi}^{H} \widehat{\bm{H}}_{bs} \bm{w}^{(n)} \bm{w}^{H} \widehat{\bm{H}}_{bs}^H \bm{\psi} \right\}-\bm{\psi}^{H} \widehat{\bm{H}}_{bs} \bm{w}^{(n)} \bm{w}^{(n),H}\widehat{\bm{H}}_{bs}^H \bm{\psi}$, $\bm{a}=$ $\text{vec}\left( \bm{\psi}\bm{\psi}^{H} \widehat{\bm{H}}_{bs} \bm{w}^{(n)} \bm{w}^H \right)
+\text{vec}\left( \bm{\psi}\bm{\psi}^{H} \widehat{\bm{H}}_{bs} \bm{w} \bm{w}^{(n),H} \right) -$ 
$\text{vec}\left( \bm{\psi}\bm{\psi}^{H} \widehat{\bm{H}}_{bs} \bm{w}^{(n)} \bm{w}^{(n),H} \right)$.
Then, we can transform  $\text{C2}$ as
\begin{align}
\label{tranC2}
\nonumber&\text{vec}^T{(\Delta \bm{H}_{bs})} \bm{A} \text{vec}{(\Delta \bm{H}_{bs} ^*)} + 2\text{Re}\left\{ \bm{a}^T  \text{vec}{(\Delta \bm{H}_{bs} ^*)} \right\} + a  \\&\geq \gamma_\text{pmin} \sigma^2, ~||\Delta \bm{H}_{bs}||_F^2 \le \xi_{bs}^2.
\end{align}
Next, we address the CSI uncertainty in \eqref{tranC2}  by applying  the General S-Procedure \cite{ssp}. In particular, we introduce the following parameters:
\begin{align*}
P=1, ~\widehat{\bm{x}} = \text{vec}{(\Delta \bm{H}_{bs}^*)}, ~\bm Q_1 = -\bm {I}_{MK}, ~\bm{g}_1 = \bm{0},  \\p_1 = \xi_{bs}^2, ~\bm Q_0 = \bm {A},~ \bm{g}_0 = \bm {a} , ~p_0 = a - \gamma_\text{pmin} \sigma^2.
\end{align*}
Accordingly, \eqref{tranC2} is equivalently transformed as
\begin{align*}
&\tilde{\text{C2}}:~  \left[\begin{array}{cc}
      \bm {A} + \widetilde{\omega}_{bs} \bm {I}_{MK}& \bm {a}  \\
       \bm {a}^T & a - \gamma_\text{pmin} \sigma^2  - \widetilde{\omega}_{bs} \xi_{bs}^2
 \end{array} \right] \succeq \bm{0},
	\end{align*}
where $\widetilde{\omega}_{bs} \geq 0$ is a slack variable and needs further optimization. Similarly,  we can approximate  C6 as
\begin{align}
\label{C6C6}
\Delta \bm{h}_{u}^H \bm{J} \Delta \bm{h}_u + 2\text{Re}\left\{ \bm{j}^H \Delta \bm{h}_u   \right\} + j  \geq \varepsilon_u, ~||\Delta \bm{h}_{u}||_2^2 \le \xi_{u}^2,
\end{align}
where
$\bm{J} = \bm{w}\bm{w}^{(n),H}  + \bm{w}^{(n)}\bm{w}^{H}  -  \bm{w}^{(n)}\bm{w}^{(n),H}$, $\bm{j} =   \bm{w}^{(n)} \bm{w}^H \widehat{\bm{h}}_u +   \bm{w} \bm{w}^{(n),H} \widehat{\bm{h}}_u  - \bm{w}^{(n)} \bm{w}^{(n),H} \widehat{\bm{h}}_u$, $j= $ $2\text{Re}\left\{  \widehat{\bm{h}}_{u}^H \bm{w}^{(n)} \bm{w}^{H} \widehat{\bm{h}}_{u} \right\} -  \widehat{\bm{h}}_{u}^H \bm{w}^{(n)} \bm{w}^{(n),H} \widehat{\bm{h}}_{u}$.
Based on this, \eqref{C6C6} is equivalently transformed as
\begin{align*}
&\tilde{\text{C6}}:~  \left[\begin{array}{cc}
      \bm {J} + \widetilde{\omega}_{u} \bm {I}_{K}& \bm {j}  \\
       \bm {j}^T & j - \varepsilon_u  - \widetilde{\omega}_{u} \xi_{u}^2
 \end{array} \right] \succeq \bm{0},
	\end{align*}
where $\widetilde{\omega}_{u} \geq 0$ is a slack variable, $\widehat{\bm{x}} = \Delta \bm{h}_u$, $P = 1$, $\bm Q_1 = -\bm {I}_{K}$, $\bm{g}_1 = \bm{0}$, $p_1 = \xi_{u}^2$, $\bm Q_0 = \bm {J}$, $\bm{g}_0 = \bm {j}$, and $p_0 = j - \varepsilon_u$.
 
The General Sign-Definiteness Principle \cite{lemmatwo} can be further leveraged to deal with the CSI uncertainty in C7. We first rewrite C7 equivalently as $\Big[\begin{array}{cc}
      \varepsilon_H & g^H \\
      g &1
 \end{array} \Big] \succeq \bm{0}$,
where $g = {(\bm{\psi}^H \bm{H}_{bs} \bm{w})^H}$. By substituting $\bm{H}_{bs} = \widehat{\bm{H}}_{bs} +\Delta \bm{H}_{bs}$ into $g$, we have the following inequality
\begin{align}
   \nonumber \left[\begin{array}{cc}
      \varepsilon_H & \widehat{g}^H \\
      \widehat{g} &1
 \end{array} \right] \succeq &-
 \left[\begin{array}{cc}
      \bm{0}_{(1 \times K)} \\
      \bm{w}^H
 \end{array} \right] \Delta \bm{H}_{bs}^H \left[\begin{array}{cc}
      \bm{\psi} & \bm{0}_{(M \times 1)}
 \end{array} \right]  \\& -  \left[\begin{array}{cc}
      \bm{\psi}^H \\
      \bm{0}_{(1 \times M)}
 \end{array} \right] \Delta \bm{H}_{bs} \left[\begin{array}{cc}
      \bm{0}_{(K \times 1)} & \bm{w}
 \end{array} \right],\nonumber
\end{align}
where $\widehat{g} = {(\bm{\psi}^H \widehat{\bm{H}}_{bs} \bm{w})^H}$. Then, according to the General Sign-Definiteness Principle, we can derive the equivalent form of  C7 as \cite{cascaded}
\begin{align*}
&\tilde{\text{C7}}:~  \left[\begin{array}{ccc}
      \varepsilon_H - b_1 M &\widehat{g}^H &\bm{0}_{(1 \times K)}\\
      \widehat{g} &1 &\xi_{bs}\bm{w}^H\\
      \bm{0}_{(K \times 1)} &\xi_{bs} \bm{w} &b_1 \bm{I}_K
 \end{array} \right] \succeq \bm{0},
	\end{align*}
where  $b_1 \geq 0$ is a slack variable,  $\bm{B} = \Big[\begin{array}{cc}
      \varepsilon_H &\widehat{g}^H \\
      \widehat{g} &1
 \end{array} \Big]$, $\bm{U}=-[\begin{array}{cc}
      \bm{0} &\bm{w}
 \end{array} ]$, $\bm{V} = [\begin{array}{cc}
      \bm{\psi} &\bm{0}
 \end{array}]$, and $\bm{X} = \Delta \bm{H}_{bs}^H$.

 By updating the objective function and constraints, $\textbf{P2.1}$ can be recast as a convex optimization problem:
		\begin{equation}\tag{$\textbf{P2.2}$}
		\begin{aligned}
			\max_{\bm{w}, \widetilde{\omega}_{bs}, \widetilde{\omega}_{u}, \varepsilon_u, \varepsilon_H, b_1} ~~&  \log_2(\varepsilon_u + \varepsilon_H + \sigma^2) - \frac{\varepsilon_H - \varepsilon_H^{(n)}}{(\ln2)(\sigma^2 + \varepsilon_H^{(n)})}\\& - \log_2(\varepsilon_H^{(n)} +\sigma^2) \\
			\text{s.t.}~~ & \text{C1},~\tilde{\text{C2}},~\tilde{\text{C6}},~\tilde{\text{C7}}, \\
            &\text{C8:}~  \widetilde{\omega}_{bs} \geq 0,  \widetilde{\omega}_{u} \geq 0,~\text{C9}: b_1\ge 0.
		\end{aligned}
	\end{equation}
The near-optimal beamforming vector $\bm{w}^{\star}$  can be derived by  solving \textbf{P2.2} at the convergence of the SCA method.

\subsection{Passive Beamforming Optimization}
	\label{OpTheda}
With the fixed $\bm{p}$ and $\bm{w}^{\star}$ obtained in Section \ref{OpTB}, we optimize  $\bm{\psi}$ by solving the sub-problem formulated as follows
\begin{equation}\tag{$\textbf{P3}$}
		\begin{aligned}
			\max_{\bm{\psi},  \varepsilon_H, b_1} ~~&  \log_2(\varepsilon_u^* + \varepsilon_H + \sigma^2) - \frac{\varepsilon_H - \varepsilon_H^{(r)}}{(\ln2)(\sigma^2 + \varepsilon_H^{(r)})} \\&- \log_2(\varepsilon_H^{(r)} +\sigma^2) \\
			\text{s.t.}~~ & \text{C2},~ \text{C3},~\text{C7},
		\end{aligned}
	\end{equation}
where $r$ denotes the iteration index for using the SCA method as shown in Section \ref{OpTB}, and $\varepsilon_u^*$ is the solution attained from \textbf{P2.2}. The solving of \textbf{P3} is also difficult due to the non-convexity of the constraints \text{C2}, \text{C3}, and \text{C7}. First, to make $\text{C3}$ tractable, a binary variable $c_{i,m} \in \{0, 1\}$, $i \in \{ 1, \ldots,   \overline\kappa  \}$ is introduced, where  $c_{i,m} = 1$ indicates that assigning the phase shift of the $m$-th element with the $i$-th parameter of $\left\{0,\frac{2\pi}{\overline\kappa}, \ldots, \frac{2\pi(\overline\kappa - 1)}{\overline\kappa}\right\}$ and $c_{i,m} = 0$ signifies that the $i$-th parameter is not selected. Therefore,  $\text{C3}$ can be expressed as
\begin{align*}
&\text{C3a}:~ \sum_{i=1}^{\overline\kappa} c_{i,m} \leq 1, \forall m,
~\text{C3b}:~\psi_m = e^{j \sum_{i=1}^{\overline\kappa} c_{i,m} f_{i}}, \forall m,
	\end{align*}
where $f_{i}$ is the $i$-th parameter of $\left\{0,\frac{2\pi}{\overline\kappa}, \ldots, \frac{2\pi(\overline\kappa - 1)}{\overline\kappa}\right\}$. $\text{C3a}$ indicates that
 each RIS element can only be assigned with one phase shift value. Nevertheless, the binary $c_{i,m}$ still poses difficulties, prompting us to reformulate it as follows
\begin{align*}
&\text{C3c}:~ c_{i,m} - c_{i,m}^2 \leq 0, \forall i,m,~~
\text{C3d}:~ 0 \leq c_{i,m} \leq 1, \forall i,m.
	\end{align*}
Then, to further eliminate the non-convexity of \text{C3c}, we implement the following  transformation
\begin{align}
  \overline{\text{C3c}}:~  c_{i,m} - (c_{i,m}^{(r)})^2 - 2c_{i,m}^{(r)} (c_{i,m} - c_{i,m}^{(r)}) \leq 0, \forall i,m.
		\end{align}
Handling  C2 here is similar to that in Section~\ref{OpTB}. Specifically, we apply the General S-Procedure \cite{ssp} to recast C2 as
\begin{align*}
&\overline{\text{C2}}:~  \left[\begin{array}{cc}
      \overline{\bm {A}} + \widetilde{\omega}_{cs} \bm {I}_{MK}& \overline{\bm {a}}  \\
       \overline{\bm {a}}^T & \overline{a} - \gamma_\text{pmin} \sigma^2  - \widetilde{\omega}_{cs} \xi_{bs}^2
 \end{array} \right] \succeq \bm{0},
	\end{align*}
where $\widetilde{\omega}_{cs} \geq 0$, $\overline{a}= 2\text{Re} \{ \bm{\psi}^{(r),H} \widehat{\bm{H}}_{bs} \bm{w}^{\star} \bm{w}^{\star,H} \widehat{\bm{H}}_{bs}^H \bm{\psi} \} -$ $ \bm{\psi}^{(r),H} \widehat{\bm{H}}_{bs} \bm{w}^{\star} \bm{w}^{\star,H} \widehat{\bm{H}}_{bs}^H \bm{\psi}^{(r)}$, $\overline{\bm{A}} = \bm{w}^{\star}\bm{w}^{\star,H} \otimes \bm{\psi}^*\bm{\psi}^{(r),T} + \bm{w}^{\star}\bm{w}^{\star,H} \otimes \bm{\psi}^{(r),*}\bm{\psi}^{T}- \left( \bm{w}^{\star}\bm{w}^{\star,H} \otimes \bm{\psi}^{(r),*}\bm{\psi}^{(r),T} \right)$, $\overline{\bm{a}} = $
$\text{vec}\left( \bm{\psi}\bm{\psi}^{(r),H} \widehat{\bm{H}}_{bs} \bm{w}^{\star} \bm{w}^{\star,H} \right) + \text{vec}\left( \bm{\psi}^{(r)}\bm{\psi}^{H} \widehat{\bm{H}}_{bs} \bm{w}^{\star} \bm{w}^{\star,H} \right) - \text{vec}\left( \bm{\psi}^{(r)}\bm{\psi}^{(r),H} \widehat{\bm{H}}_{bs} \bm{w}^{\star} \bm{w}^{\star,H} \right)$.

Similarly,  C7 can be recast as $\tilde{\text{C7}}$ as shown in Section \ref{OpTB}. We observe that only the $2 \times 2$ submatrix located in the upper left corner of $\tilde{\text{C7}}$ is controlled by $\bm{\psi}$. Thus, we can reduce the dimensionality of $\tilde{\text{C7}}$ from $(K+2) \times (K+2)$ to $2 \times 2$ as
\begin{align*}
&\overline{\text{C7}}:~  \left[\begin{array}{cc}
      \varepsilon_H - b_1 M &\widehat{g}^H \\
      \widehat{g} &1
 \end{array} \right] \succeq \bm{0}.
	\end{align*}
Till this point, problem \textbf{P3} can be reformulated as follows:
\begin{equation}\tag{$\textbf{P3.1}$}
		\begin{aligned}
			\max_{\bm{\psi},\widetilde{\omega}_{cs}, 
   \varepsilon_H, c_{i,m}, b_1} ~~&  \log_2(\varepsilon_u^* + \varepsilon_H + \sigma^2) - \frac{\varepsilon_H - \varepsilon_H^{(r)}}{(\ln2)(\sigma^2 + \varepsilon_H^{(r)})} \\&- \log_2(\varepsilon_H^{(r)} +\sigma^2) \\
			\text{s.t.}~~ & \overline{\text{C2}},~ \text{C3a},~\text{C3b},~\overline{\text{C3c}},~\text{C3d},~\overline{\text{C7}},\\
            &~\text{C9},~\text{C10:}~\widetilde{\omega}_{cs} \geq 0.
		\end{aligned}
	\end{equation}
 Similarly, the near-optimal passive beamforming, denoted by $\bm{\psi}^{\star}$, is attained by iteratively solving \textbf{P3.1}.
\subsection{MA Positions Optimization}
\label{OpMA}
After the procedures in Sections \ref{OpTB} and \ref{OpTheda}, the optimization of the MA positions is formulated as follows:
\begin{equation}\tag{$\textbf{P4}$}
		\begin{aligned}
			\max_{\bm{p}}  &   \left[\begin{array}{c}
   \mathop{\min}\limits_{\Delta{\bm{H}_{bs}}, \Delta{\bm{h}_u}}
     \{  \log_2(1+\gamma_{psr,s})  \}
 \end{array} \right] 
			~~\text{s.t.} & \text{C4},~\text{C5}.
		\end{aligned}
	\end{equation}
To handle the challenge caused by the vast solution space of \textbf{P4}, i.e., $[x_\text{min},x_\text{max}]^K \times [y_\text{min}, y_\text{max}]^K \times [z_\text{min},z_\text{max}]^K$, and ensure the computational complexity at a reasonable level, we adopt the SA-PSO algorithm \cite{HPSO} as an effective approach. The details of using the SA-PSO algorithm are summarized in Algorithm \ref{AlgorithmA} and introduced as follows:

\subsubsection{Initializing particles} 
We first generate $S$ particles, and let $\bm{p}_s^{(0)} = [\bm{p}_{s,1}^{(0)}, \bm{p}_{s,2}^{(0)},\ldots, \bm{p}_{s,K}^{(0)}]$ be the initial position of the $s$-th particle, where $\bm{p}_{s,k}^{(0)}= [ x_{s,k}^{(0)}, y_{s,k}^{(0)} , z_{s,k}^{(0)} ]$ follows $x_\text{min} \le x_{s,k}^{(0)} \le x_\text{max}$,  $y_\text{min} \le y_{s,k}^{(0)} \le y_\text{max}$ and $z_\text{min} \le z_{s,k}^{(0)} \le z_\text{max}$. Note that the position of each particle represents a possible solution to \textbf{P4}. The SA-PSO algorithm can search the particle with the best position, resulting in the best system performance. To achieve this goal, we introduce an initial velocity of each particle, denoted by $\bm{v}_s^{(0)}$,  to guarantee the update of the particles' positions.
	\subsubsection{Defining the fitness function} We define a system performance measurement function, called as fitness function, to assess the impact of updating the particles' positions on system performance, which is formulated as
	\begin{align}
	\label{FitFun}
\mathcal{F}(\bm{w}^\star,\bm{\psi}^\star,\bm{p}_s^{(q)}) = \widehat{R}_{psr}(\bm{w}^\star,\bm{\psi}^\star) - r_1 | \mathcal{R}(\bm{p}_s^{(q)}) |.
	\end{align}
 Note that a higher fitness value indicates that the current particle’s position is closer to the optimal solution.
In \eqref{FitFun}, $\widehat{R}_{psr}(\bm{w}^\star,\bm{\psi}^\star)$ represents the maximum primary rate with the solutions derived in Section \ref{OpTB} and Section \ref{OpTheda}. Let $\bm{p}_s^{(q)}$  denote the position of the $s$-th particle at the $q$-th iteration, and let $\mathcal{R}(\bm{p}_s^{(q)})$ be a set  indicating whether the MA positions are in violation of the constraint C5, which is defined as
	\begin{align}
 \mathcal{R}(\bm{p}_s^{(q)}) = \{(\bm{p}_k, \bm{p}_o) | \left \| \bm{p}_k-\bm{p}_o \right\|_2 < d_\text{min}, 1\le k \le o \le K \}.\nonumber
	\end{align}
  $| \mathcal{R}(\bm{p}_s^{(q)})|$ represents the cardinality of $\mathcal{R}(\bm{p}_s^{(q)})$. $r_1$ is a penalty factor and usually set large enough to obey $r_1 \geq \widehat{R}_{psr}(\bm{w}^\star,\bm{\psi}^\star)$.  $| \mathcal{R}(\bm{p}_s^{(q)}) |$ will tend to be zero as the number of iterations increases, making the constraint C5 be satisfied.

	\subsubsection{Updating the positions and  velocities of particles} This procedure  is implemented in the step 6 of Algorithm \ref{AlgorithmA}. In particular, we update
	the position of the $s$-th particle based on its  locally best position, i.e., $\bm{p}_{s,\text{best}}^{(q)}$,  and the globally optimal position determined by all particles, i.e., $\bm{p}_{\text{best}}^{(q)}$. To describe this updating procedure, we introduce an inertia weight parameter $\omega$, two step factors, denoted by $c_1$ and $c_2$, and two random parameters $r_2$ and $r_3$ uniformly generated within $[0,1]$. The updates of positions and velocities are respectively given by
	\begin{align}
	\label{UpdateV}
\nonumber\bm{v}_s^{(q+1)} &= \omega \bm{v}_s^{(q)} +c_1 r_2 (\bm{p}_{s,\text{best}}^{(q)} -\bm{p}_s^{(q)} ) + c_2 r_3 (\bm{p}_{\text{best}}^{(q)} -\bm{p}_s^{(q)} ),\\& s= 1,\ldots, S,\\
\label{UpdateP}
\bm{p}_s^{(q+1)} &= \bm{p}_s^{(q)} + \bm{v}_s^{(q+1)}, ~s= 1,\ldots, S.
	\end{align}
 Furthermore, we define $h_{s,k}^{(q)} \in \{ x_{s,k}^{(q)}, y_{s,k}^{(q)}, z_{s,k}^{(q)} \}$, which is utilized to limit the updated positions violating the constraint C5 by the following operation
 \begin{equation}
 	  	\label{CheckP}
 	  	h_{s,k}^{(q)}  = \left \{
 	  	\begin{aligned}
 	  	       &h_\text{min},  ~~~&\text{if}~~ h_{s,k}^{(q)} < h_\text{min},\\
             & h_\text{max}, ~~~ &\text{if}~~ h_{s,k}^{(q)} > h_\text{max},\\
              & h_{s,k}^{(q)},  ~&\text{otherwise},
 	  	\end{aligned}
 	  	\right.
 	  \end{equation}
 	   where $h_\text{min}$ and $h_\text{max}$ represent the minimum and maximum values of the specified range, respectively.

 	  \subsubsection{Updating the globally optimal position}\label{SA}
The steps 13-23 of Algorithm \ref{AlgorithmA} summarize the procedure for updating $\bm{p}_\text{best}^{(q)}$.
 	   Note that the traditional PSO algorithm is easy to be limited to a local optimal solution. To overcome this deficiency, we utilize the SA method leveraging a greedy strategy with a probability $\epsilon$ to accept solutions worse than the current one. As the value of $\epsilon$ significantly affects the convergence performance and solution accuracy, we update $\epsilon$ as follows~\cite{HPSO}:
 	 \begin{align}
 	 \label{Probability}
            \epsilon = \exp \left(\frac{\mathcal{F} (\bm{w}^\star, \bm{\psi}^\star,  \bm{p}_{\text{best}}^{(q)} )- \mathcal{F} (\bm{w}^\star, \bm{\psi}^\star,  \bm{p}_{\text{best}}^{(q+1)} )  } {T^{(q)}} \right).
 	 \end{align}
 	 In \eqref{Probability}, $T^{(q)}$ represents the system temperature for the SA method, which is updated as $T^{(q+1)} = \frac{Q-q}{Q} T^{(q)}$, with $Q$ being the maximum iteration time. It is worth noting that the initialization of the system temperature, i.e., $T^{(0)}$, is empirical.

 	  To update $\bm{p}_\text{best}^{(q)}$, we  first sort $\{\bm{p}_s^{(q+1)}\}_{s=1}^S$ in decreasing order based on their fitness function values in step 13, and then decide whether to adopt the greedy strategy and accept a worse solution $\bm{p}_{\wr}^{(q+1)}$ in steps 14-23 to extend the solution diversity, where $\wr$ is a random value generated from $[0, q+1]$. This procedure is the key of the SA-PSO algorithm to achieve better performance than the PSO algorithm.

  	 \begin{algorithm}
  	\caption{SA-PSO algorithm for updating MA positions.}
  	\label{AlgorithmA}
  	\begin{algorithmic}[1]
  		\STATE {Initialization: The iteration number $q=0$, the maximum iteration time $Q$, $\bm{p}_s^{(q)}$, $\bm{v}_s^{(q)}$,   and $T^{(q)}$.}
  		\STATE{Obtain $\mathcal{F} (\bm{w}^\star,\bm{\psi}^\star,\bm{p}_s^{(q)})$ according to \eqref{FitFun}.}
  		\STATE{Let  $\bm{p}_{s,\text{best}}^{(q)} = \bm{p}_s^{(q)}$ and $\bm{p}_{\text{best}}^{(q)} = \arg \max \{ \mathcal{F} (\bm{w}^\star,\bm{\psi}^\star,\bm{p}_1^{(q)}), \ldots, \mathcal{F} (\bm{w}^\star,\bm{\psi}^\star,\bm{p}_S^{(q)}) \}$.}
  		\WHILE{$q < Q$} 
  		\FOR{$s=1:S$}
  			\STATE {Update $\bm{v}_s^{(q+1)}$  and $\bm{p}_s^{(q+1)}$
     based on \eqref{UpdateV}-\eqref{CheckP}.}
            \IF{$\mathcal{F} (\bm{w}^\star,\bm{\psi}^\star,\bm{p}_s^{(q+1)}) > \mathcal{F} (\bm{w}^\star,\bm{\psi}^\star,\bm{p}_{s,\text{best}}^{(q)}) $}
                \STATE{$\bm{p}_{s,\text{best}}^{(q+1)} = \bm{p}_s^{(q+1)}$.}
            \ELSE{}
                 \STATE{$\bm{p}_{s,\text{best}}^{(q+1)} = \bm{p}_{s,\text{best}}^{(q)}$.}
            \ENDIF
         \ENDFOR
  		
                \STATE{Sort $\bm{p}_s^{(q+1)}$  in decreasing order by $\mathcal{F} (\bm{w}^\star,\bm{\psi}^\star,\bm{p}_s^{(q+1)})$ for $s\in \{1,\ldots,S\}$, and let $\bm{p}_\text{temp} =\bm{p}_1^{(q+1)}$.}
                \IF{$\mathcal{F}(\bm{w}^\star,\bm{\psi}^\star,\bm{p}_\text{temp}) < \mathcal{F}(\bm{w}^\star,\bm{\psi}^\star,\bm{p}_\text{best}^{(q)})$  }
                    \STATE{Update $\epsilon$ according to \eqref{Probability}.}
                    	\IF{$\epsilon > \text{rand}(0,1)$}
  		                      \STATE{$\bm{p}_\text{best}^{(q+1)} = \bm{p}_{\wr}^{(q+1)}$. }
  		                 \ELSE{}
  		                      \STATE{$\bm{p}_\text{best}^{(q+1)} = \bm{p}_\text{best}^{(q)}$.}
  		                 \ENDIF
                    \ELSE{}
                    \STATE{$ \bm{p}_\text{best}^{(q+1)} = \bm{p}_\text{temp}$.  }
                \ENDIF
  		  \STATE{$q=q+1$.}
  		\ENDWHILE
  	    \STATE{Return $\bm{p}$.}
  	\end{algorithmic}
  \end{algorithm}

\subsection{Summary and Analysis of the Proposed AO Algorithm}
We summarize the details presented above and outline the solving procedure for \textbf{P1} in Algorithm \ref{AlgorithmB}.  {We first anlayze the convergence of Algorithm \ref{AlgorithmB} theoretically, which is determined by the updating trend of the objective function value of \textbf{P1}. It is obvious that the objective function value of \textbf{P2.2} (i.e., \textbf{P1}) is non-decreasing after implementing the SCA method  to find its near-optimal solution from step 5 to step 8. It is because compared to the solution in the previous AO iteration, the obtained near-optimal solution in this AO iteration can always result in a non-worse objective function value. 
Similarly, the objective function value of \textbf{P3.1} (\textbf{P1}) is also non-decresing after finding its near-optimal solution from step 11 to step 14. Moreover, the SA-PSO shown in Algorithm \ref{AlgorithmA} can ensure that  the fitness value  based on the updated globally best MA positions has a non-decreasing trend. Thus, we conclude that the objective function value of \textbf{P1} is always non-decreasing  after sequentially solving the three sub-problems in each AO iteration, indicating that Algorithm \ref{AlgorithmB} will converge. The numerical results indicated in Fig. \ref{fig:shoulian} will also verify this analysis.}

The complexity associated with Algorithm \ref{AlgorithmB} is expressed as $\mathcal{O} (  A_1SQ + A_1A_2( MK+2K+4  )^{\frac{1}{2}} A_4  [ A_4^2 +   A_4( MK+1 )^2 + A_4(  K+1 )^2  + A_4( K+2 )^2    +  ( MK+1)^3  + ( K+1 )^3  +( K+2 )^3  ]  +A_1A_3 (MK+3  )^{\frac{1}{2}} A_5  [ A_5^2 +  A_5( MK+1 )^2 +4A_5  + (MK+1 )^3 +8  ] )$ \cite{cascaded}, where $A_1$ represents the iteration number  required to achieve convergence of Algorithm \ref{AlgorithmB}, $A_2$ and $A_3$ are the iteration numbers required to execute the SCA method for \textbf{P2.2} and \textbf{P3.1}, respectively, $A_4$ and $A_5$ are the amounts of variables for \textbf{P2.2} and \textbf{P3.1}, respectively.
  	 \begin{algorithm}
  	\caption{ The algorithm for solving \textbf{P1}.}
  	\label{AlgorithmB}
  	\begin{algorithmic}[1]

  		\STATE {Initialization:  The  MA positions $\bm{p}^{(0)}$, the tolerance value $\varrho$,  and the AO iteration index $\varsigma=0$.}

  		\REPEAT
           \STATE{$\varsigma=\varsigma+1$.}
           \STATE{Initialization:  $n=0$ and $\bm{w}^{(n)}$.}
           \REPEAT
                \STATE{$n=n+1$.}
                \STATE{ Solve \textbf{P2.2} to update $\bm{w}^{(n)}$. }
           \UNTIL{the SCA method converges.}

           \STATE{Let $\bm{w}^{(\star,\varsigma)} = \bm{w}^{(n)}$.}
           \STATE{Initialization:  $r=0$ and  $\bm{\psi}^{(r)}$.}

            \REPEAT
                \STATE{$r=r+1$.}
                \STATE{ Solve \textbf{P3.1} to update $\bm{\psi}^{(r)}$.}
           \UNTIL{the SCA method converges.}
           \STATE{Let  $\bm{\psi}^{(\star,\varsigma)} = \bm{\psi}^{(r)}$.}

           \STATE{Update $\bm{p}^{(\varsigma)}$ by implementing Algorithm \ref{AlgorithmA}.}
    \UNTIL{the AO framework converges. }
       \STATE{Return $\bm{w}^{\star}=\bm{\psi}^{(\star,\varsigma)}$, $\bm{\psi}^{\star}= \bm{\psi}^{(\star,\varsigma)}$, and $\bm{p}^{\star}= \bm{p}^{(\varsigma)}$.}
  	\end{algorithmic}
  \end{algorithm}

\section{Robust Transmission Design for  CSR}
\label{sicsr}
This section presents the robust transmission design for the CSR scenario. Similar to \textbf{P1}, the optimization problem for the CSR scenario is expressed as
\begin{equation}\tag{$\textbf{P5}$}
		\begin{aligned}
			\max_{\bm{w}, \bm{\psi}, \bm{p}} ~~& \left[\begin{array}{c}
   \mathop{\min}\limits_{\Delta{\bm{H}_{bs}}, \Delta{\bm{h}_u}}
     \{   R_{csr}  \}
 \end{array} \right] \\
                \text{s.t.}~~ & \text{C1},~\text{C3},~\text{C4},~\text{C5}, \\
			& \text{C11:~} L |\bm{\psi}^H \bm{H}_{bs} \bm{w}|^2 \geq  \gamma_\text{cmin} \sigma^2, ~||\Delta \bm{H}_{bs}||_F \le \xi_{bs}.
		\end{aligned}
\end{equation}
{The objective function of \textbf{P5} is also formulated as a max-min form to achieve the primary rate maximization for the CSR scenario under the worst-case within the channel uncertainty region.} In \textbf{P5}, C11 is the QoS constraint on the secondary transmission for the CSR scenario, and $\gamma_\text{cmin}$ is the associated SNR threshold. Note that the main difference between \textbf{P1} and \textbf{P5} lies in the objective function, and the one of \textbf{P5} is more complicated. Similar to \textbf{P1}, the AO framework is leveraged to cope with the variable coupling. The detailed procedures are shown below.
\subsection{Transmit Beamforming Optimization}
	\label{OpcsrTB}
The optimization of $\bm{w}$ derived from \textbf{P5} is formulated as
\begin{equation}\tag{$\textbf{P6}$}
		\begin{aligned}
			\max_{\bm{w}}  & \left[\begin{array}{c}
   \mathop{\min}\limits_{\Delta{\bm{H}_{bs}}, \Delta{\bm{h}_u}}
     \{  R_{csr}  \}
 \end{array} \right] ~
			\text{s.t.}  & \text{C1},~\text{C11}.\\
            \end{aligned}
	    \end{equation}
The objective function in \textbf{P5} can be rewritten as
\begin{align}
\label{inequalitycsr}
  \nonumber&\mathop{\min}\limits_{\Delta{\bm{H}_{bs}}, \Delta{\bm{h}_u}}
     \{  R_{csr}  \}
      \\&\nonumber=   \frac{1}{2} \log_2 (1+ \frac{\mathop{\min}\limits_{\Delta{\bm{H}_{bs}}, \Delta{\bm{h}_u}} \left\{ |\left( \bm{h}_{u}^H + \bm{\psi}^H \bm{H}_{bs} \right) \bm{w}  |^2\right\}}{\sigma^2} ) \\&+ \frac{1}{2} \log_2 (1+ \frac{\mathop{\min}\limits_{\Delta{\bm{H}_{bs}}, \Delta{\bm{h}_u}} \left\{ |\left( \bm{h}_{u}^H - \bm{\psi}^H \bm{H}_{bs} \right) \bm{w}  |^2\right\}}{\sigma^2} ).
\end{align}
Thus, \textbf{P6} can be equivalently  reformulated as follows:
\begin{equation}\tag{$\textbf{P6.1}$}
		\begin{aligned}
			\max_{\bm{w}, \epsilon_u,\epsilon_H } ~~&   \frac{1}{2} \log_2(1+\frac{\epsilon_u}{\sigma^2}) + \frac{1}{2} \log_2(1+\frac{\epsilon_H}{\sigma^2})
  \\
			\text{s.t.}~~ & \text{C1},~{\text{C11}}, \\
                &\text{C12}:~ \epsilon_u \leq \mathop{\min}\limits_{ \Delta{\bm{H}_{bs}}, \Delta{\bm{h}_u}} \left\{ |\left( \bm{h}_{u}^H + \bm{\psi}^H \bm{H}_{bs} \right) \bm{w}  |^2 \right\},\\
                &\text{C13}:~ \epsilon_H \leq \mathop{\min}\limits_{ \Delta{\bm{H}_{bs}}, \Delta{\bm{h}_u}} \left\{ |\left( \bm{h}_{u}^H - \bm{\psi}^H \bm{H}_{bs} \right) \bm{w}  |^2 \right\},
		\end{aligned}
	\end{equation}
where $\epsilon_u$ and $\epsilon_H$ are positive variables.   To solve \textbf{P6.1}, we need to  transform ${\text{C11}}$ into a tractable form. Similar to ${\text{C2}}$, 
introducing a slack variable $\widehat{\omega}_{bs} \geq 0$, ${\text{C11}}$ can be rewritten as
\begin{align*}
&\tilde{\text{C11}}:~  \left[\begin{array}{cc}
      \bm {A} + \widehat{\omega}_{bs} \bm {I}_{MK}& \bm {a}  \\
       \bm {a}^T & a - \gamma_\text{cmin} \sigma^2 /L  - \widehat{\omega}_{bs} \xi_{bs}^2
 \end{array} \right] \succeq \bm{0}.
	\end{align*}
 Then,  by applying the SCA and the General S-Procedure \cite{ssp}, we  further transform C12 and C13 into convex constraints. Specifically,  by substituting $\bm{H}_{bs} = \widehat{\bm{H}}_{bs} + \Delta \bm{H}_{bs}$ and $\bm{h}_{u} = \widehat{\bm{h}}_{u} + \Delta \bm{h}_{u}$ into C12, we have $| [ {(  \widehat{\bm{h}}_u + \Delta \bm{h}_{u} )}^H + \bm{\psi}^H (  \widehat{\bm{H}}_{bs} + \Delta \bm{H}_{bs}  ) ] \bm{w}|^2$, which is lower bounded by $\bm{x}^H \widehat{\bm{A}} \bm{x}+ 2\text{Re}\left\{ \widehat{\bm{a}}^H  \bm{x} \right\} + \widehat{a}$,
where $\widehat{\bm{A}} = \bm{D} + \bm{D}^H -\bm{Z}$, $
\widehat{\bm{a}} = \bm{d}_1 + \bm{d}_2 - \tilde{\bm{z}}$,
 $\widehat{a} = 2\text{Re}\left\{ d \right\} - z$,
\begin{align*}
&\tilde{\bm{z}} = \left[\begin{array}{cc}
     \bm{w}^{(n)} \bm{w}^{(n),H} \left( \widehat{\bm{h}}_u +  \widehat{\bm{H}}_{bs}^H \bm{\psi} \right) \\
       \text{vec}^*  \left( \bm{\psi} \left(  \widehat{\bm{h}}_u^H + \bm{\psi}^{H} \widehat{\bm{H}}_{bs}    \right)  \bm{w}^{(n)} \bm{w}^{(n),H} \right)
 \end{array} \right],\\&\bm{x} = \left[\begin{array}{cc}
       \Delta \bm{h}_{u}^H ~~\text{vec}^H  \left(\Delta \bm{H}_{bs}^*  \right)
 \end{array} \right]^H,\\
&\bm{D} = \left[\begin{array}{cc}
     \bm{w}^{(n)} \\
       \bm{w}^{(n)} \otimes \bm{\psi}^{*}
 \end{array} \right] \left[\begin{array}{cc}
     \bm{w}^H ~~ \bm{w}^H \otimes \bm{\psi}^T
 \end{array} \right],\\&
\bm{Z} = \left[\begin{array}{cc}
     \bm{w}^{(n)} \\
       \bm{w}^{(n)} \otimes \bm{\psi}^{*}
 \end{array} \right] \left[\begin{array}{cc}
     \bm{w}^{(n),H}~~ \bm{w}^{(n),H}\otimes\bm{\psi}^{T}
 \end{array} \right],\\
&\bm{d}_1 = \left[\begin{array}{cc}
     \bm{w} \bm{w}^{(n),H} \left( \widehat{\bm{h}}_u +  \widehat{\bm{H}}_{bs}^H \bm{\psi} \right) \\
       \text{vec}^*  \left( \bm{\psi} \left(  \widehat{\bm{h}}_u^H + \bm{\psi}^{H} \widehat{\bm{H}}_{bs}    \right)  \bm{w}^{(n)} \bm{w}^H \right)
 \end{array} \right],\\&
\bm{d}_2 = \left[\begin{array}{cc}
     \bm{w}^{(n)} \bm{w}^{H} \left( \widehat{\bm{h}}_u +  \widehat{\bm{H}}_{bs}^H \bm{\psi} \right) \\
       \text{vec}^*  \left( \bm{\psi} \left(  \widehat{\bm{h}}_u^H + \bm{\psi}^{H} \widehat{\bm{H}}_{bs}    \right)  \bm{w} \bm{w}^{(n),H} \right)
 \end{array} \right],
\\
&d =  \left(  \widehat{\bm{h}}_u^H + \bm{\psi}^{H} \widehat{\bm{H}}_{bs}    \right)  \bm{w}^{(n)} \bm{w}^{H}  \left(  \widehat{\bm{h}}_u + \widehat{\bm{H}}_{bs}^H \bm{\psi}    \right),\\&
z = \left(  \widehat{\bm{h}}_u^H + \bm{\psi}^{H} \widehat{\bm{H}}_{bs}    \right)  \bm{w}^{(n)} \bm{w}^{(n),H}  \left(  \widehat{\bm{h}}_u + \widehat{\bm{H}}_{bs}^H \bm{\psi}    \right).
\end{align*}
Then, C12 can be approximately rewritten as follows
\begin{align}
\nonumber&\bm{x}^H \widehat{\bm{A}} \bm{x}+ 2\text{Re}\left\{ \widehat{\bm{a}}^H  \bm{x} \right\} + \widehat{a} \geq \epsilon_u,\\&||\Delta \bm{H}_{bs}||_F^2 \le \xi_{bs}^2,~||\Delta \bm{h}_{u}||_2^2 \le \xi_{u}^2.
\end{align}
Next, we formulate the CSI uncertainties in quadratic forms:
\begin{align*}
\bm{x}^H \left[\begin{array}{cc}
       \bm{I}_K &\bm{0} \\
       \bm{0} & \bm{0}
 \end{array} \right] \bm{x}  \leq  \xi_{u}^2 \text{ and }
 \bm{x}^H \left[\begin{array}{cc}
       \bm{0} &\bm{0} \\
       \bm{0} & \bm{I}_{MK}
 \end{array} \right] \bm{x}  \leq \xi_{bs}^2.
\end{align*}
Finally, slack variables $\overline{\omega}_{bs} \geq 0$ and $\overline{\omega}_{u} \geq 0$ are introduced, and  C12 is converted as
\begin{align*}
&\tilde{\text{C12}}:~  \left[\begin{array}{cc}
      \widehat{\bm {A}} +\left[\begin{array}{cc} \overline{\omega}_{u} \bm {I}_{K} & \bm{0}\\ \bm{0} & \overline{\omega}_{bs} \bm{I}_{MK}  \end{array} \right]
      & \widehat{\bm {a}}  \\
       \widehat{\bm {a}}^H & C_{full}
 \end{array} \right] \succeq \bm{0},
	\end{align*}
where $C_{full} = \widehat{a} -\epsilon_u - \overline{\omega}_{u} \xi_{u}^2 - \overline{\omega}_{bs} \xi_{bs}^2$. Similarly,  C13 can be approximated as
\begin{align*}
&\tilde{\text{C13}}:~  \left[\begin{array}{cc}
      \widehat{\bm {A_c}} +\left[\begin{array}{cc} \overline{\omega}_{cu} \bm {I}_{K} & \bm{0}\\ \bm{0} & \overline{\omega}_{cb} \bm{I}_{MK}  \end{array} \right]
      & \widehat{\bm {a_c}}  \\
       \widehat{\bm {a_c}}^H & \widehat{C}_{full}
 \end{array} \right] \succeq \bm{0},
	\end{align*}
 where $\overline{\omega}_{cu} \geq 0$,  $\overline{\omega}_{cb} \geq 0$, $\widehat{C}_{full} = \widehat{a_c} -\epsilon_H - \overline{\omega}_{cu} \xi_{u}^2 - \overline{\omega}_{cb} \xi_{bs}^2$, $\widehat{\bm{A_c}} = \bm{D_c} + \bm{D_c}^H -\bm{Z_c}$,
  $\widehat{\bm{a_c}} = \bm{d_{c1}} + \bm{d_{c2}} - \tilde{\bm{z_c}}$, $\widehat{a_c} = 2\text{Re}\left\{ d_c \right\} - z_c$,
\begin{align*}
&\bm{D_c} = \left[\begin{array}{cc}
     \bm{w}^{(n)} \\
       -\bm{w}^{(n)} \otimes \bm{\psi}^{*}
 \end{array} \right] \left[\begin{array}{cc}
     \bm{w}^H ~~ -\bm{w}^H \otimes \bm{\psi}^T
 \end{array} \right], \\
&\bm{Z_c} = \left[\begin{array}{cc}
     \bm{w}^{(n)} \\
       -\bm{w}^{(n)} \otimes \bm{\psi}^{*}
 \end{array} \right] \left[\begin{array}{cc}
     \bm{w}^{(n),H}~~ -\bm{w}^{(n),H}\otimes\bm{\psi}^{T}
 \end{array} \right],\\
&\bm{d_{c1}} = \left[\begin{array}{cc}
     \bm{w} \bm{w}^{(n),H} \left( \widehat{\bm{h}}_u -  \widehat{\bm{H}}_{bs}^H \bm{\psi} \right) \\
       -\text{vec}^*  \left( \bm{\psi} \left(  \widehat{\bm{h}}_u^H - \bm{\psi}^{H} \widehat{\bm{H}}_{bs}    \right)  \bm{w}^{(n)} \bm{w}^H \right)
 \end{array} \right],\\
&\bm{d_{c2}} = \left[\begin{array}{cc}
     \bm{w}^{(n)} \bm{w}^{H} \left( \widehat{\bm{h}}_u -  \widehat{\bm{H}}_{bs}^H \bm{\psi} \right) \\
       -\text{vec}^*  \left( \bm{\psi} \left(  \widehat{\bm{h}}_u^H - \bm{\psi}^{H} \widehat{\bm{H}}_{bs}    \right)  \bm{w} \bm{w}^{(n),H} \right)
 \end{array} \right],\\
&\tilde{\bm{z_c}} = \left[\begin{array}{cc}
     \bm{w}^{(n)} \bm{w}^{(n),H} \left( \widehat{\bm{h}}_u -  \widehat{\bm{H}}_{bs}^H \bm{\psi} \right) \\
       -\text{vec}^*  \left( \bm{\psi} \left(  \widehat{\bm{h}}_u^H - \bm{\psi}^{H} \widehat{\bm{H}}_{bs}    \right)  \bm{w}^{(n)} \bm{w}^{(n),H} \right)
 \end{array} \right],\\
&d_c =  \left(  \widehat{\bm{h}}_u^H - \bm{\psi}^{H} \widehat{\bm{H}}_{bs}    \right)  \bm{w}^{(n)} \bm{w}^{H}  \left(  \widehat{\bm{h}}_u - \widehat{\bm{H}}_{bs}^H \bm{\psi}    \right),\\
&z_c = \left(  \widehat{\bm{h}}_u^H - \bm{\psi}^{H} \widehat{\bm{H}}_{bs}    \right)  \bm{w}^{(n)} \bm{w}^{(n),H}  \left(  \widehat{\bm{h}}_u - \widehat{\bm{H}}_{bs}^H \bm{\psi}    \right).
\end{align*}
 With the updated constraints, \textbf{P6.1} can be reformulated as the following convex optimization problem
		\begin{equation}\tag{$\textbf{P6.2}$}
		\begin{aligned}
			\max_{\bm{w}, \epsilon_u,\epsilon_H, {\omega}_{w} } ~~&  \frac{1}{2}  \log_2(1 +\frac{\epsilon_u}{\sigma^2}) + \frac{1}{2}  \log_2(1 +\frac{\epsilon_H}{\sigma^2})\\
			\text{s.t.}~~ & \text{C1},~\tilde{\text{C11}},~\tilde{\text{C12}},~\tilde{\text{C13}},\\
   &\text{C14}:~ \widehat{\omega}_{bs} \geq 0,\\
   &\text{C15}:~\overline{\omega}_{bs} \geq 0,~\overline{\omega}_{u} \geq 0,\\
   &\text{C16}:~\overline{\omega}_{cb} \geq 0,~\overline{\omega}_{cu} \geq 0,
		\end{aligned}
	\end{equation}
 where ${\omega}_{w} = \left\{\overline{\omega}_{bs}, \overline{\omega}_{u},\widehat{\omega}_{bs},\overline{\omega}_{cb},\overline{\omega}_{cu}\right\}$. \textbf{P6.2} can be solved by using convex optimization approaches \cite{CVX}.

\subsection{Passive Beamforming Optimization}
	\label{OpcsrTheda}
Considering the sub-problem of optimizing $\bm{\psi}$, the constraints \text{C3a}, \text{C3b}, $\overline{\text{C3c}}$ and \text{C3d} can be substituted for the original constraint C3 as the descriptions in Section \ref{OpTheda}. After that, this sub-problem is formulated as
\begin{equation}\tag{$\textbf{P7}$}
		\begin{aligned}
			\max_{\bm{\psi}, \epsilon_u,\epsilon_H,  c_{i,m} } ~~&   \frac{1}{2} \log_2(1+\frac{\epsilon_u}{ \sigma^2})   + \frac{1}{2} \log_2(1+\frac{\epsilon_H}{ \sigma^2})
  \\
			\text{s.t.}~~ & \text{C3a}, ~\text{C3b}, ~\overline{\text{C3c}}, ~\text{C3d}, ~\text{C11},~\text{C12},~\text{C13}.                                \end{aligned}
	\end{equation}
To solve \textbf{P7}, further transformations with respect to the constraints C11, C12 and C13 are required.  Specifically, similar to C2, we introduce a slack variable $\widehat{\omega}_{cs} \geq 0$ and utilize the General S-Procedure to transform  $\text{C11}$  into $\overline{\text{C11}}$  as:
\begin{align*}
&\overline{\text{C11}}:~  \left[\begin{array}{cc}
      \overline{\bm {A}} + \widehat{\omega}_{cs} \bm {I}_{MK}& \overline{\bm {a}}  \\
       \overline{\bm {a}}^T & \overline{a} - \gamma_\text{cmin} \sigma^2/L  - \widehat{\omega}_{cs} \xi_{bs}^2
 \end{array} \right] \succeq \bm{0}.
	\end{align*}
Then, we introduce slack variables $\ddot{\omega}_{u} \geq 0$, $\ddot{\omega}_{bs} \geq 0$, $\ddot{\omega}_{cu} \geq 0$, $\ddot{\omega}_{cb} \geq 0$, and respectively  transform $\text{C12}$ and $\text{C13}$ as
\begin{align*}
&\overline{\text{C12}}:~  \left[\begin{array}{cc}
      \ddot{\bm {A}} +\left[\begin{array}{cc} \ddot{\omega}_{u} \bm {I}_{K} & \bm{0}\\ \bm{0} & \ddot{\omega}_{bs} \bm{I}_{MK}  \end{array} \right]
      & \ddot{\bm {a}}  \\
       \ddot{\bm {a}}^H & \overline{C}_{full}
 \end{array} \right] \succeq \bm{0},\\
 &\overline{\text{C13}}:~  \left[\begin{array}{cc}
      \ddot{\bm {A_c}} +\left[\begin{array}{cc} \ddot{\omega}_{cu} \bm {I}_{K} & \bm{0}\\ \bm{0} & \ddot{\omega}_{cb} \bm{I}_{MK}  \end{array} \right]
      & \ddot{\bm {a_c}}  \\
       \ddot{\bm {a_c}}^H & \ddot{C}_{full}
 \end{array} \right] \succeq \bm{0},
	\end{align*}
 where $\overline{C}_{full} = \ddot{a} -\epsilon_u - \ddot{\omega}_{u} \xi_{u}^2 - \ddot{\omega}_{bs} \xi_{bs}^2$, $\ddot{C}_{full} = \ddot{a_c} -\epsilon_H-$ $\ddot{\omega}_{cu} \xi_{u}^2 - \ddot{\omega}_{cb} \xi_{bs}^2$, $\ddot{\bm{A}} = \ddot{\bm{D}} + \ddot{\bm{D}}^H -\ddot{\bm{Z}}$, $\ddot{\bm{A_c}} = \ddot{\bm{D_c}} + \ddot{\bm{D_c}}^H -\ddot{\bm{Z_c}}$, $\ddot{\bm{a}} = \ddot{\bm{d}}_1 + \ddot{\bm{d}}_2 - \ddot{\bm{z}}$, $\ddot{\bm{a_c}} = \ddot{\bm{d_c}}_1 + \ddot{\bm{d_c}}_2 - \ddot{\bm{z_c}}$, $\ddot{a} = 2\text{Re}\left\{ \ddot{d} \right\} - \ddot{z}$, $\ddot{a_c} = 2\text{Re}\left\{ \ddot{d_c} \right\} - \ddot{z_c}$,  $\bm{x} = \left[\begin{array}{cc}
       \Delta \bm{h}_{u}^H ~~\text{vec}^H  \left(\Delta \bm{H}_{bs}^*  \right)
 \end{array} \right]^H$,  
 \begin{align*}
 &\ddot{D} = \left[\begin{array}{cc}
     \bm{w} \\
       \bm{w} \otimes \bm{\psi}^{(r),*}
 \end{array} \right] \left[\begin{array}{cc}
     \bm{w}^H ~~\bm{w}^H \otimes \bm{\psi}^T
 \end{array} \right],\\
 &\ddot{\bm{Z}} = \left[\begin{array}{cc}
     \bm{w}\\
       \bm{w} \otimes \bm{\psi}^{(r),*}
 \end{array} \right] \left[\begin{array}{cc}
     \bm{w}^{H}~~ \bm{w}^{H} \otimes \bm{\psi}^{(r),T}
 \end{array} \right],\\
 &\ddot{D_c} = \left[\begin{array}{cc}
     \bm{w} \\
      - \bm{w} \otimes \bm{\psi}^{(r),*}
 \end{array} \right] \left[\begin{array}{cc}
     \bm{w}^H ~~ -\bm{w}^H \otimes \bm{\psi}^T
 \end{array} \right],\\
 &\ddot{\bm{Z_c}} = \left[\begin{array}{cc}
     \bm{w}\\
       -\bm{w} \otimes \bm{\psi}^{(r),*}
 \end{array} \right] \left[\begin{array}{cc}
     \bm{w}^{H}~~ -\bm{w}^{H} \otimes \bm{\psi}^{(r),T}
 \end{array} \right],\\
 &\ddot{\bm{d}}_1 = \left[\begin{array}{cc}
     \bm{w} \bm{w}^{H} \left( \widehat{\bm{h}}_u +  \widehat{\bm{H}}_{bs}^H \bm{\psi}^{(r)} \right) \\
       \text{vec}^*  \left( \bm{\psi} \left(  \widehat{\bm{h}}_u^H + \bm{\psi}^{(r),H} \widehat{\bm{H}}_{bs}    \right)  \bm{w} \bm{w}^H \right)
 \end{array} \right],\\
 &\ddot{\bm{d}}_2 = \left[\begin{array}{cc}
     \bm{w} \bm{w}^{H} \left( \widehat{\bm{h}}_u +  \widehat{\bm{H}}_{bs}^H \bm{\psi} \right) \\
       \text{vec}^*  \left( \bm{\psi}^{(r)} \left(  \widehat{\bm{h}}_u^H + \bm{\psi}^{H} \widehat{\bm{H}}_{bs}    \right)  \bm{w} \bm{w}^{H} \right)
 \end{array} \right],\\
 &\ddot{\bm{d_c}}_1 = \left[\begin{array}{cc}
     \bm{w} \bm{w}^{H} \left( \widehat{\bm{h}}_u -  \widehat{\bm{H}}_{bs}^H \bm{\psi}^{(r)} \right) \\
       -\text{vec}^*  \left( \bm{\psi} \left(  \widehat{\bm{h}}_u^H - \bm{\psi}^{(r),H} \widehat{\bm{H}}_{bs}    \right)  \bm{w} \bm{w}^H \right)
 \end{array} \right],
 \end{align*}
 \begin{align*}
 &\ddot{\bm{d_c}}_2 = \left[\begin{array}{cc}
     \bm{w} \bm{w}^{H} \left( \widehat{\bm{h}}_u -  \widehat{\bm{H}}_{bs}^H \bm{\psi} \right) \\
       -\text{vec}^*  \left( \bm{\psi}^{(r)} \left(  \widehat{\bm{h}}_u^H - \bm{\psi}^{H} \widehat{\bm{H}}_{bs}    \right)  \bm{w} \bm{w}^{H} \right)
 \end{array} \right],\\
 &\ddot{\bm{z}} = \left[\begin{array}{cc}
     \bm{w} \bm{w}^{H} \left( \widehat{\bm{h}}_u +  \widehat{\bm{H}}_{bs}^H \bm{\psi}^{(r)} \right) \\
       \text{vec}^*  \left( \bm{\psi}^{(r)} \left(  \widehat{\bm{h}}_u^H + \bm{\psi}^{(r),H} \widehat{\bm{H}}_{bs}    \right)  \bm{w} \bm{w}^{H} \right)
 \end{array} \right],\\
 &\ddot{\bm{z_c}} = \left[\begin{array}{cc}
     \bm{w} \bm{w}^{H} \left( \widehat{\bm{h}}_u -  \widehat{\bm{H}}_{bs}^H \bm{\psi}^{(r)} \right) \\
       -\text{vec}^*  \left( \bm{\psi}^{(r)} \left(  \widehat{\bm{h}}_u^H - \bm{\psi}^{(r),H} \widehat{\bm{H}}_{bs}    \right)  \bm{w} \bm{w}^{H} \right)
 \end{array} \right],\\
 &\ddot{d} =  \left(  \widehat{\bm{h}}_u^H + \bm{\psi}^{(r),H} \widehat{\bm{H}}_{bs}    \right)  \bm{w}\bm{w}^{H}  \left(  \widehat{\bm{h}}_u + \widehat{\bm{H}}_{bs}^H \bm{\psi}    \right),\\
 &\ddot{d_c} =  \left(  \widehat{\bm{h}}_u^H - \bm{\psi}^{(r),H} \widehat{\bm{H}}_{bs}    \right)  \bm{w}\bm{w}^{H}  \left(  \widehat{\bm{h}}_u - \widehat{\bm{H}}_{bs}^H \bm{\psi}    \right),\\
 &\ddot{z} = \left(  \widehat{\bm{h}}_u^H + \bm{\psi}^{(r),H} \widehat{\bm{H}}_{bs}    \right)  \bm{w} \bm{w}^{H}  \left(  \widehat{\bm{h}}_u + \widehat{\bm{H}}_{bs}^H \bm{\psi}^{(r)}    \right),\\
 &\ddot{z_c} = \left(  \widehat{\bm{h}}_u^H - \bm{\psi}^{(r),H} \widehat{\bm{H}}_{bs}    \right)  \bm{w} \bm{w}^{H}  \left(  \widehat{\bm{h}}_u - \widehat{\bm{H}}_{bs}^H \bm{\psi}^{(r)}    \right).
   \end{align*}
To this end, we find the passive beamforming solution for the CSR scenario by solving \textbf{P7.1}
\begin{equation}\tag{$\textbf{P7.1}$}
		\begin{aligned}
			\max_{\bm{\psi}, \epsilon_u,  \epsilon_H,\widehat{\omega}_{w}, c_{i,m} } ~~& \frac{1}{2} \log_2(1 +\frac{\epsilon_u}{\sigma^2}) + \frac{1}{2} \log_2(1 +\frac{\epsilon_H}{\sigma^2})\\
			\text{s.t.}~~ & \text{C3a},~\text{C3b},~\overline{\text{C3c}},~\text{C3d},\overline{~\text{C11}},~\overline{\text{C12}},~\overline{\text{C13}},\\
            &~\text{C17}:~\widehat{\omega}_{cs} \geq 0, ~\ddot{\omega}_{bs} \geq 0,~\ddot{\omega}_{u} \geq 0,\\
            &~\text{C18}: ~\ddot{\omega}_{cb} \geq 0,~\ddot{\omega}_{cu} \geq 0,
		\end{aligned}
	\end{equation}
where $\widehat{\omega}_{w} = \left\{ \ddot{\omega}_{bs},\ddot{\omega}_{u},\widehat{\omega}_{cs},\ddot{\omega}_{cb},\ddot{\omega}_{cu} \right\}$.

\subsection{MA Positions Optimization}
\label{OpcsrMA}
With the attained solutions in Sections \ref{OpcsrTB} and \ref{OpcsrTheda}, we continue to search the optimal MA positions by addressing \textbf{P8}, formulated as
\begin{equation}\tag{$\textbf{P8}$}
		\begin{aligned}
			\max_{\bm{p}} ~ &   \left[\begin{array}{c}
   \mathop{\min}\limits_{\Delta{\bm{H}_{bs}}, \Delta{\bm{h}_u}}
     \{  \log_2(1+\gamma_{csr,s})  \}
 \end{array} \right] ~
			\text{s.t.}  & \text{C4},~\text{C5}.
		\end{aligned}
	\end{equation}
We also solve \textbf{P8} by using the SA-PSO algorithm. Similarly, the fitness function for the CSR scenario is defined as
	\begin{align}
	\label{FitFuncsr}
\widehat{\mathcal{F}}(\bm{w}^\star,\bm{\psi}^\star,\bm{p}_s^{(q)}) = \widehat{R}_{csr}(\bm{w}^\star,\bm{\psi}^\star) - r_4 \left| \mathcal{R}(\bm{p}_s^{(q)})\right|,
	\end{align}
where $\widehat{R}_{csr}(\bm{w}^\star,\bm{\psi}^\star)$ is the maximum primary rate with the fixed positions of MAs, $r_4$ is a penalty factor satisfying  $\widehat{R}_{csr}(\bm{w}^\star,\bm{\psi}^\star) -r_4 \le 0$. The other steps of utilizing the SA-PSO algorithm here are similar to those described in Section \ref{OpMA} and neglected to avoid duplication.

The whole procedure to solve \textbf{P5} is also similar to Algorithm \ref{AlgorithmB} and neglected for simplicity. The associated  complexity of solving \textbf{P5} is $\mathcal{O} ( \hat{A}_1SQ + \hat{A}_1 A_6( 3MK+2K+3)^{\frac{1}{2}} A_8 [ A_8^2 +   A_8( MK+1 )^2 + 2A_8(MK+K+1)^2  + ( MK+1)^3 + 2( MK+K+1 )^3  ]  +\hat{A}_1A_7 (3MK+2K+3)^{\frac{1}{2}} A_9  [ A_9^2 +   A_9( MK+1 )^2 +2A_9(MK+K+1)^2  + (MK+1 )^3 + 2( MK+K+1 )^3  ] )$, where $\hat{A}_1$ represents the iteration number  required to achieve convergence of solving \textbf{P5},
$A_6$ and $A_7$ represent the iteration numbers required to execute the SCA method of solving \textbf{P6.2} and \textbf{P7.1}, respectively, $A_8$ and $A_9$ are the amounts of variables for \textbf{P6.2} and \textbf{P7.1}, respectively.

\section{Simulation Results}
\label{results}
This section presents numerical results to demonstrate the  efficacy of the proposed AO framework and show the superior performance of the proposed MA empowered SR scheme. A three-dimensional topology is used to model the simulation environment, with the position of the  RIS setting at  (0 m, 30 m, 40 m). We model the moving region of the MAs as a square area on the \emph{x-y} plane of size $[-\frac{A}{2},\frac{A}{2}]\times [-\frac{A}{2},\frac{A}{2}]$ with the center coordinate (3 m, 0 m, 0 m) and $z_k = $ 0 m, where $A = 3\lambda$ is the side length, $\lambda$ is the carrier wavelength with $\lambda = 0.1$ m. The AoAs and AoDs are uniformly generated from $[-\frac{\pi}{2},\frac{\pi}{2}]$. We set the numbers of transmit and receive paths to the same, i.e., $L_p^t = L_s^t = L_{\kappa}^t = \overline{L} = 9$. The path-response matrices $\bm{\Sigma}_{b}$, $\bm{\Sigma}_{s}$ and $\bm{\Sigma}_{u}$ are expressed as $\text{diag}\{[\widehat{m}_{b,1}, \ldots, \widehat{m}_{b,\overline{L}}]\}$, $\text{diag}\{[\widehat{m}_{s,1}, \ldots, \widehat{m}_{s,\overline{L}}]\}$ and $\text{diag}\{[\widehat{m}_{u,1}, \ldots, \widehat{m}_{u,\overline{L}}]\}$, respectively, where $\widehat{m}_{b,\overline{l}}$, $\widehat{m}_{s,\overline{l}}$ and $\widehat{m}_{u,\overline{l}}$  are the associated complex responses of the $\overline{l}$-th path for $\overline{l}= 1, \ldots, \overline{L}$ and all satisfy $\mathcal{CN}(0,\widehat{v} \cdot \Upsilon^{-\nu} {/} \overline{L})$.  $\Upsilon$ represents the distance between two nodes, $\widehat{v}$ represents the path-loss, and $\nu$ is the pass-loss exponent. Denote the proportion of cascaded and direct CSI uncertainties by $g_{bs} \in {[0,1)}$ and $g_u \in {[0,1)}$, respectively. Then, the associated variances of $\Delta{\bm{H}_{bs}}$ and $\Delta{\bm{h}_{u}}$ are modeled as $\xi_{bs}^2 = g_{bs}^2 ||\widehat{\bm{H}}_{bs}||_F^2$ and $\xi_{u}^2 = g_u^2 ||\widehat{\bm{h}}_u||_2^2$ \cite{Two-timescale}. 
The remaining parameters are set as follows: $K = 4$, $M = 8$, $\widehat{v} = -10$ dB, $\nu = 1.3$, $d_\text{min} = 0.5\lambda$, $L = 50$, $P_\text{max} = 38$ dBm, $\sigma^2 = 10^{-12}$, $S = 150$, $Q = 150$, $c_1 = c_2 =1.4$, $\omega = 1.2$, $r_1 = r_4 =50$, $g_{bs} = 0.05$, $g_{u} = 0.1$, $\overline \kappa = 8$, and $\varrho = 10^{-2}$.

For performance comparison, we consider the following benchmark schemes. 1) {Proposed scheme with PSO}: In this scheme, the PSO algorithm is used to update the MA positions for our proposed MA scheme. 2) {FPA scheme}: In this scheme, the antennas at the PT are with fixed-positions. In addition, the optimization of transmit and passive beamforming shown in Sections \ref{sanpsr} and \ref{sicsr} are also considered. 3) {Random passive beamforming scheme}: In this scheme, only the optimization of transmit beamforming and MA positions is considered, while the passive beamforming is randomly generated.

\begin{figure}[t]
  \centering
  \includegraphics[width=0.32\textwidth]{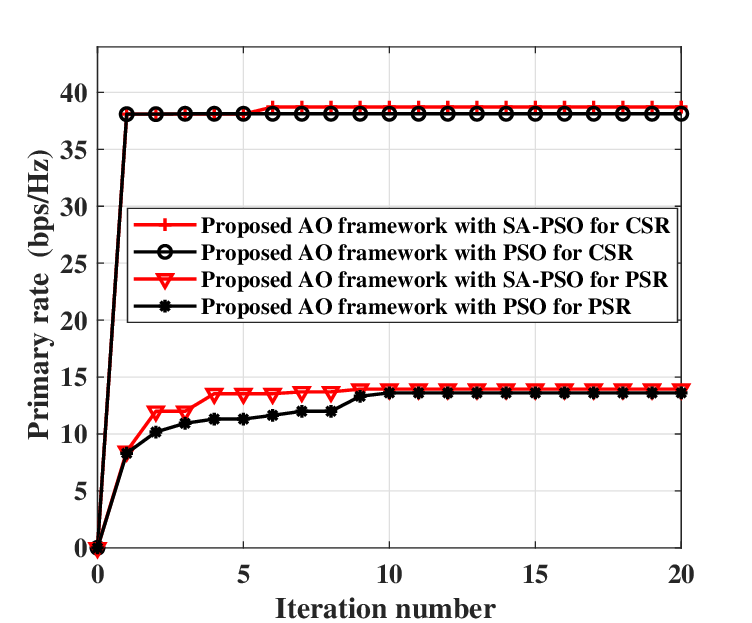}
  \caption{ Convergence performance of the proposed AO framework. }
  \label{fig:shoulian}
\end{figure}

{In the following, we first investigate the numerical results for the single-PU scenario and then study the extension to the multi-PU scenario.

\subsection{Single-PU Scenario}
We first consider the single-PU scneario with the position of the PU setting at (0, 60 m, 0 m).} Fig. \ref{fig:shoulian} describes the convergence performance of the proposed  AO framework with the SA-PSO and PSO algorithms for both the PSR and CSR scenarios. {We first observe that the primary rate of the CSR scenario is much better than that of the PSR scenario. This is due to the fact that the secondary transmission for the CSR scenario can offer a multipath to improve the delivery of primary information, but the secondary transmission for the PSR scenario causes interference.} Compared to the PSR scenario, the AO framework for the CSR  scenario has a faster convergence speed. Specifically, the AO framework for the CSR scenario with the SA-PSO algorithm  only needs 6 iterations to converge, while the AO framework for the PSR scenario with the SA-PSO algorithm  needs 9 iterations. It is mainly because the fractional expression (or difference expression) existing in  the PSR scenario makes it more challenging to search for a satisfactory solution.

\begin{figure}[t]
  \centering
  \begin{minipage}[b]{0.24\textwidth}
    \includegraphics[width=\textwidth]{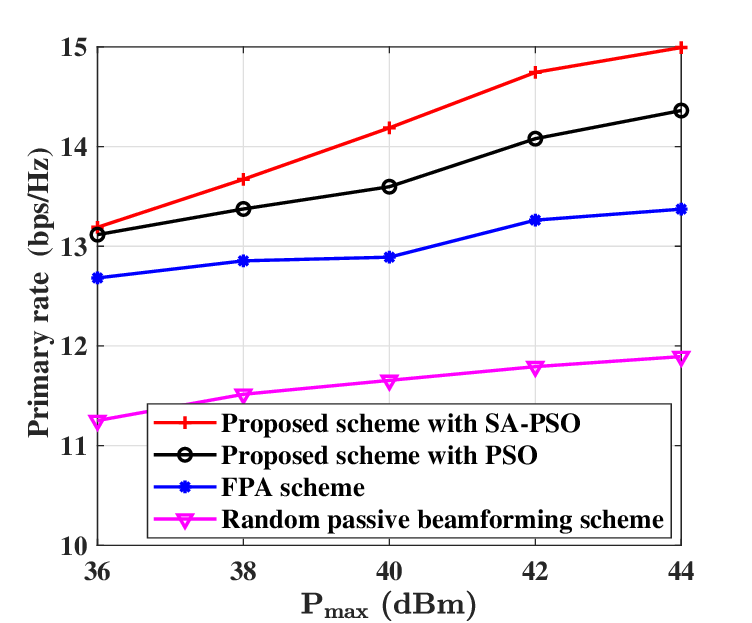}
    \caption{Primary rate versus the maximum transmit power for the PSR.}
    \label{fig:psrp}
  \end{minipage}
  \hfill
  \begin{minipage}[b]{0.24\textwidth}
    \includegraphics[width=\textwidth]{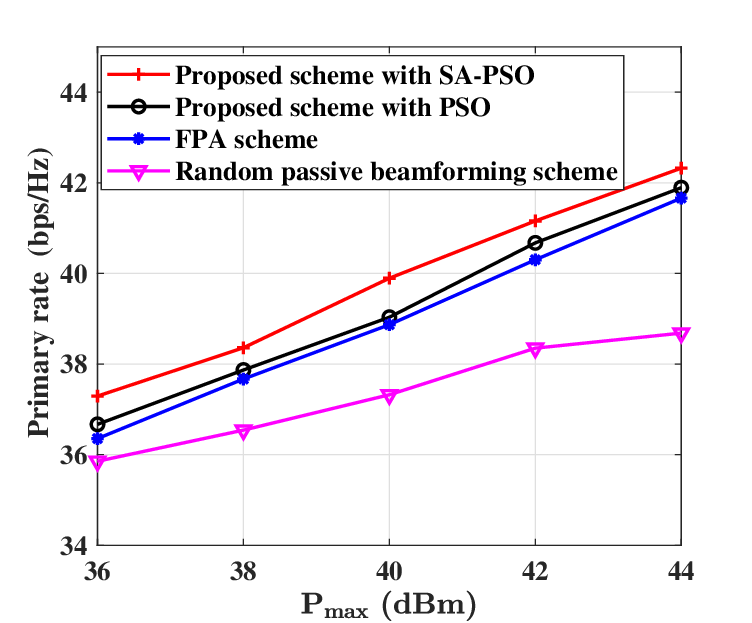}
    \caption{Primary rate versus the maximum transmit power for the CSR.}
    \label{fig:csrp}
  \end{minipage}
\end{figure}

Fig. \ref{fig:psrp} illustrates the variation in primary rate as a function of the maximum transmit power for the PSR scenario.  It is clear that setting a larger maximum transmit power improves primary rates across all schemes. The proposed scheme with the SA-PSO algorithm can ensure better performance than the proposed scheme with the PSO algorithm.
 It is because that the incorporation of the SA method can prevent the PSO algorithm from falling into local optimal solutions by adopting a greedy strategy, thus improving the solution accuracy. Fig. \ref{fig:psrp} also shows that the FPA scheme performs worse than the proposed MA scheme as it cannot move antennas to establish favorable channel conditions. Moreover, we find that the performance of the random passive beamforming scheme is always the worst, indicating that the adjustment of phase shifts is a key factor for performance improvement. 
 Fig. \ref{fig:csrp} shows how the maximum transmit power affects the primary rate for the CSR scenario. As observed in Fig. \ref{fig:csrp}, the primary rates of the proposed scheme with both SA-PSO and PSO algorithms are also significantly higher than those of the other two benchmarks. From comprehensive comparisons between Fig. \ref{fig:psrp} and Fig. \ref{fig:csrp}, we again discover that the CSR scenario performs significantly better than the PSR scenario. {This discrepancy arises from the fact that the secondary transmission for the CSR scenario is treated as a beneficial multipath  for delivering primary information, while it is regarded as interference for the PSR scenario and damages the associated primary transmission.}

\begin{figure}[t]
  \centering
  \begin{minipage}[b]{0.23\textwidth}
    \includegraphics[width=\textwidth]{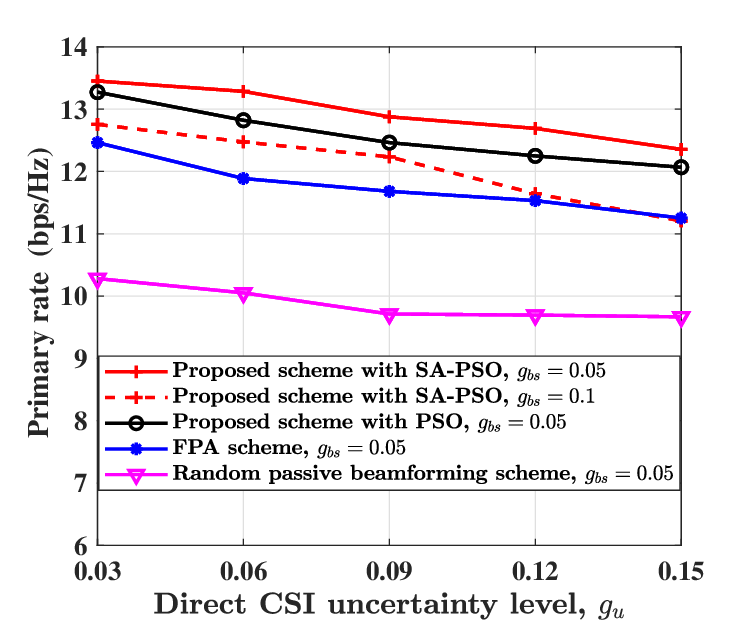}
    \caption{Primary rate versus the CSI uncertainty level for the PSR.}
    \label{fig:psru}
  \end{minipage}
  \hfill
  \begin{minipage}[b]{0.23\textwidth}
    \includegraphics[width=\textwidth]{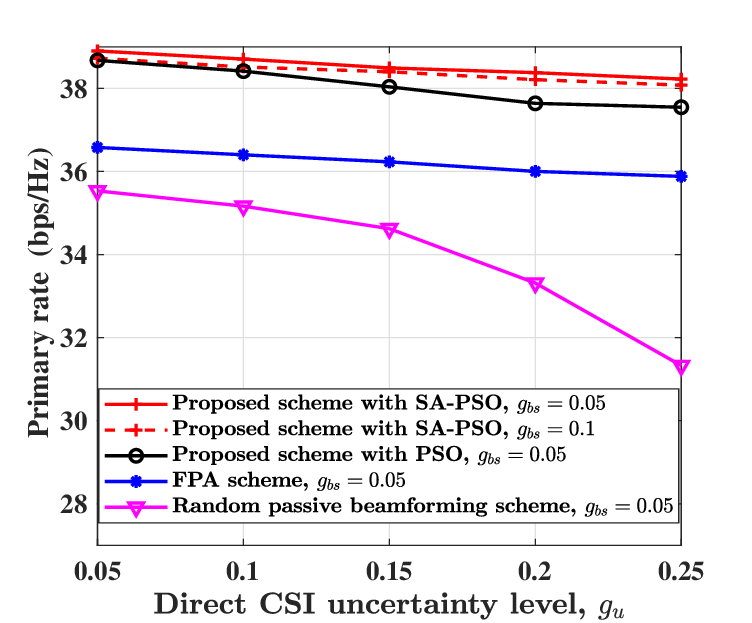}
    \caption{Primary rate versus the CSI uncertainty level for the CSR.}
    \label{fig:csru}
  \end{minipage}
\end{figure}



\begin{figure}[t]
  \centering
  \begin{minipage}[b]{0.24\textwidth}
    \includegraphics[width=\textwidth]{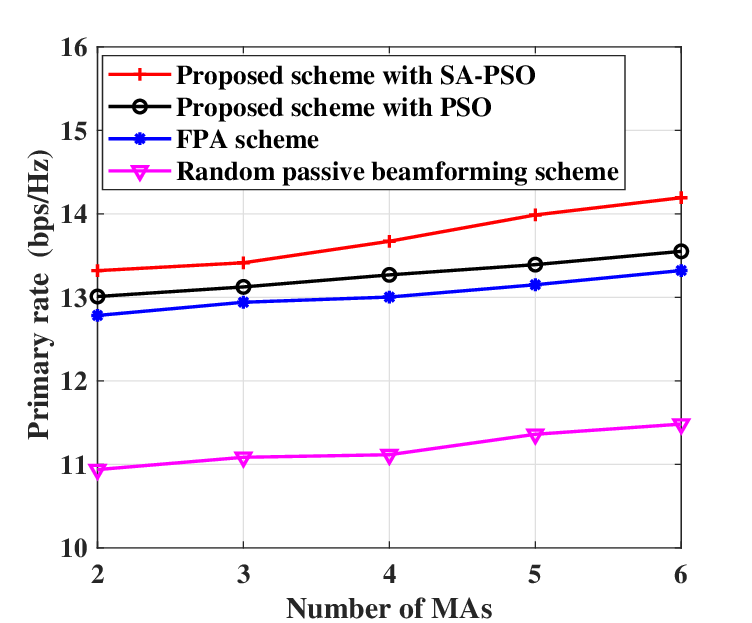}
    \caption{Primary rate versus the number of MAs for the PSR.}
    \label{fig:psrk}
  \end{minipage}
  \hfill
  \begin{minipage}[b]{0.24\textwidth}
    \includegraphics[width=\textwidth]{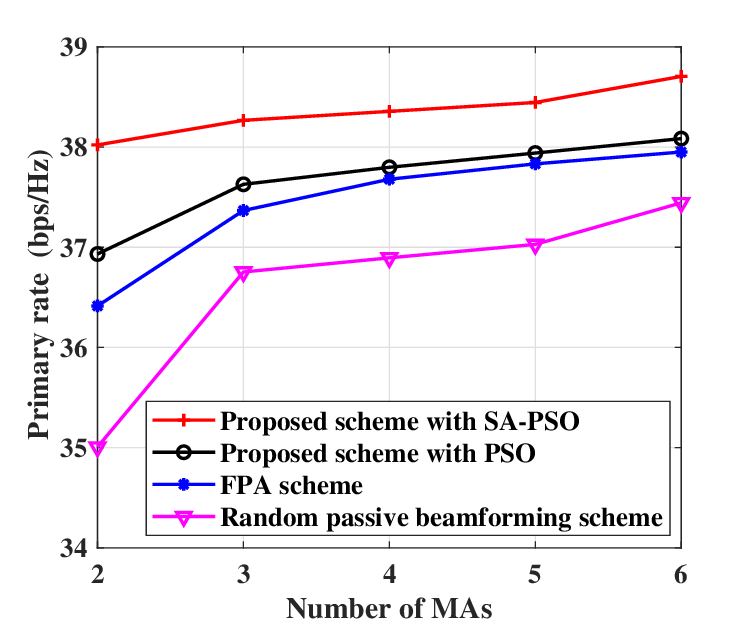}
    \caption{Primary rate versus the number of MAs for the CSR.}
    \label{fig:csrk}
  \end{minipage}
\end{figure}

{Fig. \ref{fig:psru} and Fig. \ref{fig:csru} investigate the relationship between the primary rate and the channel uncertainty levels $g_{bs}$ and $g_{u}$ for the PSR and CSR scenarios, respectively. One common observation from Fig. \ref{fig:psru} and Fig. \ref{fig:csru} is that exacerbating the channel uncertainty level of either the direct link or the cascaded link damages the primary transmission performance. It is because that by exacerbating the channel uncertainty level, the region of finding the worst-case channel condition expands, resulting in a worse primary rate.}  For example, for the PSR scenario,  varying the direct CSI uncertainty from 0.03 to 0.15, the primary rates of the proposed scheme with  $g_{bs} = 0.05$ for both algorithms reduce 1.10 bps/Hz and 1.21 bps/Hz, respectively. In Fig. \ref{fig:psru}, when $g_u = 0.15$, the FPA scheme with  $g_{bs} = 0.05$  can even attain a larger primary rate than that of the proposed scheme with the SA-PSO algorithm and $g_{bs} = 0.1$, indicating that an accuracy CSI is the basis for satisfactory system performance. However, the random passive beamforming scheme with $g_{bs}=0.05$  still performs the worst.




Fig. \ref{fig:psrk} and Fig. \ref{fig:csrk} show how the number of MAs  affects the primary rate for the PSR and CSR scenarios,  respectively. It is obvious that  higher primary rates can be attained with exploiting more MAs due to the enhanced spatial diversity. Consistent with Fig. \ref{fig:psru} and Fig. \ref{fig:csru}, the proposed scheme with the SA-PSO algorithm attains the highest primary rate. This finding underscores the efficacy of adopting the MAs in constructing  better transmission environments and  the SA method in finding accurate solutions.

\begin{figure}[t]
  \centering
  \begin{minipage}[b]{0.24\textwidth}
    \includegraphics[width=\textwidth]{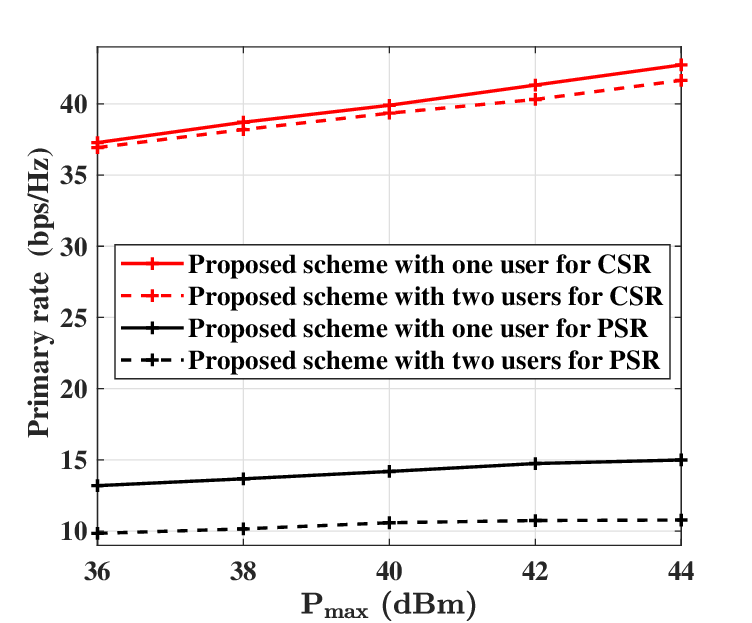}
    \caption{Primary rate versus the maximum transmit power for the scenario with different number of PUs.}
    \label{fig:usersp}
  \end{minipage}
  \hfill
  \begin{minipage}[b]{0.24\textwidth}
    \includegraphics[width=\textwidth]{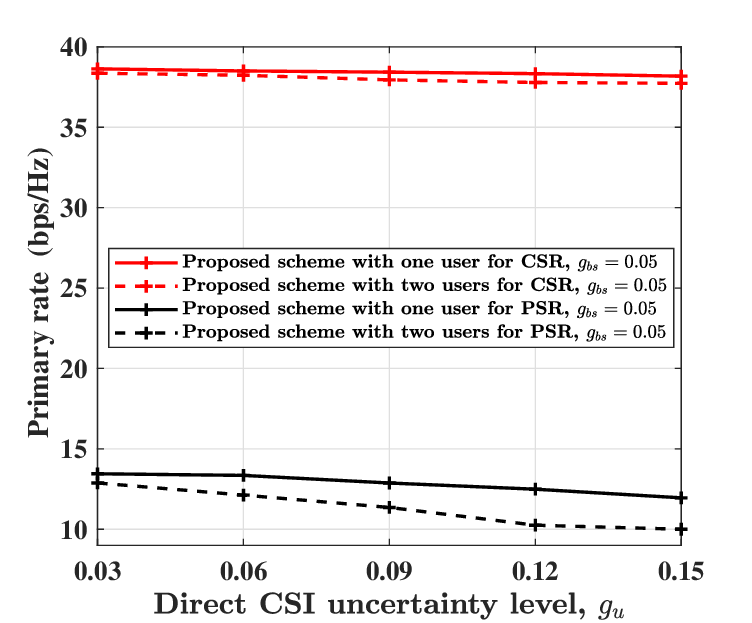}
    \caption{Primary rate versus  the CSI
uncertainty level for the scenario with different number of PUs.}
    \label{fig:usersu}
  \end{minipage}
\end{figure}

{
\subsection{Multi-PU Scenario}
\label{sub:multi_pu_scenario}
In this sub-section, we consider the scneario with multiple PUs, for which the broadcasting transmission model is considered. Specifically, the PT sends common primary signals to all the PUs, and these PUs independetly decode the received primary and secondary signals, respectively. To this end, we  formulate the primary rate maximization problem with the objective function as  
\begin{align}
\label{MultiPSR}
\max \limits_{\bm{w}, \bm{\psi}, \bm{p}} ~~[ \mathop{\min}\limits_{\Delta{\bm{H}_{bs,\varpi}}, \Delta{\bm{h}_{u,\varpi}}} 
     \left\{  \log_2(1+\gamma_{psr,s,\varpi})  \right\}_{\varpi=1}^\varPi  ]
\end{align}
      for the PSR scenario, and 
\begin{align}
\label{MultiCSR}
 \max_{\bm{w}, \bm{\psi}, \bm{p}} ~~[ 
   \mathop{\min}\limits_{\Delta{\bm{H}_{bs,\varpi}}, \Delta{\bm{h}_{u,\varpi}}}
     \left\{  \log_2(1+\gamma_{csr,s,\varpi})  \right\}_{\varpi=1}^\varPi ]
     \end{align}
     for the CSR scenario, respectively. In \eqref{MultiPSR} and \eqref{MultiCSR}, $\gamma_{psr,s,\varpi}$ and $\gamma_{csr,s,\varpi}$ are the corresponding SINR and SNR for decoding primary signals at the $\varpi$-th PU, $\Delta{\bm{H}_{bs,\varpi}}$ and $\Delta{\bm{h}_{u,\varpi}}$ are the channel uncertainties related with the $\varpi$-th PU, and $\varPi$ is the number of PUs. The constraints C2 and C11 also need to be updated to satisfy the QoS constraints on decoding secondary signals at each PU for the PSR and CSR scenarios, respectively. For the updated optimization problems, we can still utilize the proposed algorithm  shown in Section \ref{sanpsr} and Section \ref{sicsr} to find their respective solutions.

Without loss of generality, we consider a  two-PU scenario with their positions setting at (0, 60 m, 0 m) and (0, 80 m, 0 m), respectively. In Fig. \ref{fig:usersp} and Fig. \ref{fig:usersu}, we compare the primary rate performance between the single-PU scenario and the multi-PU scenario. Firstly,  Fig. \ref{fig:usersp} and Fig. \ref{fig:usersu} indicate the observation that the primary rate of the two-PU scenario is smaller than that of the single-PU scenario. It is because with the objective functions defined in \eqref{MultiPSR} and \eqref{MultiCSR}, the primary rate is determined by the PU at (0, 80 m, 0 m) with the worst-case channel uncertainty since the PU at (0, 80 m, 0 m) is farther away from the PT (or the RIS). Secondly, we find that compared to the PSR scenario, the performance gap between the single-PU scenario and the two-PU scenario for the CSR  is slighter, which shows the robustness of the CSR  to address the  CSI uncertainty level.
} 



\section{Conclusions}
\label{Conc}
In this paper, we have proposed to apply the MA technology to improve robust performance of RIS enabled SR systems. Specifically, multiple MAs have deployed at the PT to construct a favorable transmission environment in conjunction with the utilization of RIS. To combat the existence of direct and cascaded CSI uncertainties for both PSR and CSR scenarios, we have designed robust transmission schemes with optimized beamforming vectors and MA positions. Moreover, we  have validated the superiority of the proposed MA scheme over the FPA scheme through numerical results. We have also  observed that the CSR scenario can achieve a much larger primary rate than the PSR scenario since the CSR scenario ensures a mutually beneficial relationship between primary and secondary transmissions. 

{In the future work, we can extend the current work from the following directions. First, it is promising to apply  the MA technology in all the transceivers and RIS  to further boost the system performance by exploiting more spatial DoFs.  However, the above extensions may result in extremely complex optimization problems with  variables couplings and uncertain dynamics. To address this issue, optimization-driven learning algorithms  are worthy of investigations. Moreover, designing robust RIS modulation schemes is  a key perspective  against the absence of direct links and unsatisfied decoding performance for SR communications.}


\begin{thebibliography}{}
\providecommand{\url}[1]{#1}
\csname url@samestyle\endcsname
\providecommand{\newblock}{\relax}
\providecommand{\bibinfo}[2]{#2}
\providecommand{\BIBentrySTDinterwordspacing}{\spaceskip=0pt\relax}
\providecommand{\BIBentryALTinterwordstretchfactor}{4}
\providecommand{\BIBentryALTinterwordspacing}{\spaceskip=\fontdimen2\font plus
\BIBentryALTinterwordstretchfactor\fontdimen3\font minus
  \fontdimen4\font\relax}
\providecommand{\BIBforeignlanguage}[2]{{%
\expandafter\ifx\csname l@#1\endcsname\relax
\typeout{** WARNING: IEEEtran.bst: No hyphenation pattern has been}%
\typeout{** loaded for the language `#1'. Using the pattern for}%
\typeout{** the default language instead.}%
\else
\language=\csname l@#1\endcsname
\fi
#2}}
\providecommand{\BIBdecl}{\relax}
\BIBdecl

\end{thebibliography}


\begin{thebibliography}{10}
\bibliographystyle{IEEEtran}
\bibitem{MYVTC}
B. Lyu et al., ``Primary rate maximization in movable antennas empowered symbiotic radio communications,'' in \emph{Proc. IEEE 99th Veh. Technol. Conf. (VTC2024-Spring)}, Singapore, Singapore, 2024, pp. 1-6.

\bibitem{ITU}
ITU-R WP5D, ``ITU-R framework for IMT-2030,'' Jul. 2023. Available:
https://www.itu.int/en/ITU-R/study-groups/rsg5/rwp5d/imt-2030.



\bibitem{LiangSurvey}
Y. -C. Liang et al., ``Symbiotic radio: Cognitive backscattering communications for future wireless networks,'' \emph{IEEE Trans. Cognit. Commun. Netw.}, vol. 6, no. 4, pp. 1242-1255, Dec. 2020.

\bibitem{LongIoT}
R. Long, Y. -C. Liang, H. Guo, G. Yang, and R. Zhang, ``Symbiotic radio: A new communication paradigm for passive Internet of Things,'' \emph{IEEE Internet Things J.}, vol. 7, no. 2, pp. 1350-1363, Feb. 2020. 

\bibitem{Zeng}
Z. Dai, R. Li, J. Xu, Y. Zeng, and S. Jin, ``Rate-region characterization and channel estimation for cell-free symbiotic radio communications,'' \emph{IEEE Trans.  Commun.}, vol. 71, no. 2, pp. 674-687, Feb. 2023.




\bibitem{QQZhang}
Q. Zhang, Y. -C. Liang, H. -C. Yang, and H. V. Poor, ``Mutualistic mechanism in symbiotic radios: When can the primary and secondary transmissions be mutually beneficial?,'' \emph{IEEE Trans. Wireless Commun.}, vol. 21, no. 10, pp. 8036-8050, Oct. 2022.


\bibitem{ZengMassive}
J. Xu, Z. Dai, and Y. Zeng, ``MIMO symbiotic radio with massive backscatter devices: Asymptotic analysis and precoding optimization,'' \emph{IEEE Trans.  Commun.}, vol. 71, no. 9, pp. 5487-5502, Sep. 2023.




\bibitem{risone}
P. Ramezani, B. Lyu, and A. Jamalipour, ``Toward RIS-enhanced integrated terrestrial/non-terrestrial connectivity in 6G,'' \emph{IEEE Netw.}, vol. 37, no. 3, pp. 178-185, Jun. 2023.


\bibitem{HuaRIS}
M. Hua et al., ``Secure intelligent reflecting surface aided integrated sensing and communication,'' \emph{IEEE Trans. Wireless Commun.}, vol. 23, no. 1, pp. 575-591, Jan. 2024.

\bibitem{RISSurvey}
Q. Wu et al., ``Intelligent surfaces empowered wireless network: Recent advances and the road to 6G,'' \emph{Proc. IEEE}, vol. 112, no. 7, pp. 724-763, Jul. 2024.


\bibitem{MengHua}
M. Hua et al., ``A novel wireless communication paradigm for intelligent reflecting surface based symbiotic radio systems,'' \emph{IEEE Trans. Signal Process.}, vol. 70, pp. 550-565, 2022.

\bibitem{HuaUAV}

M. Hua, L. Yang, Q. Wu, C. Pan, C. Li, and A. L. Swindlehurst, ``UAV-assisted intelligent reflecting surface symbiotic radio system,'' \emph{IEEE Trans. Wireless Commun.}, vol. 20, no. 9, pp. 5769-5785, Sep. 2021.



\bibitem{Zhou}
H. Zhou, Q. Zhang, Y. -C. Liang, and Y. Pei, ``Assistance-transmission tradeoff for RIS-assisted symbiotic radios,'' \emph{IEEE Trans. Wireless Commun.}, vol. 23, no. 7, pp. 6838-6855, Jul. 2024.



\bibitem{BPSK}
Q. Zhang, Y. -C. Liang, and H. V. Poor, ``Reconfigurable intelligent surface assisted MIMO symbiotic radio networks,'' \emph{IEEE Trans. Commun.}, vol. 69, no. 7, pp. 4832-4846, July 2021.




\bibitem{LYUBIN}
B. Lyu et al., ``Robust secure transmission for active RIS enabled symbiotic radio multicast communications,'' \emph{IEEE Trans. Wireless Commun.}, vol. 22, no. 12, pp. 8766-8780, Dec. 2023.

\bibitem{YongjunRSMA}
Y. Xu et al., ``Robust beamforming and rate optimization for RIS-aided symbiotic radio systems with RSMA,'' \emph{IEEE Commun. Lett.}, vol. 28, no. 10, pp. 2328-2332, Oct. 2024.

\bibitem{Ng}
S. Hu et al., ``Robust and secure sum-rate maximization for multiuser MISO downlink systems with self-sustainable IRS,''  \emph{IEEE Trans. Commun.}, vol. 69, no. 10, pp. 7032-7049, Oct. 2021.

\bibitem{Pan}
L. Zhang, C. Pan, Y. Wang, H. Ren, and K. Wang, ``Robust beamforming design for intelligent reflecting surface aided cognitive radio systems with imperfect cascaded CSI,''  \emph{IEEE Trans. Cognit. Commun. Netw.}, vol. 8, no. 1, pp. 186-201, Mar. 2022.

\bibitem{PanTwo}
S. Hong, C. Pan, H. Ren, K. Wang, K. K. Chai, and A. Nallanathan, ``Robust transmission design for intelligent reflecting surface-aided secure communication systems with imperfect cascaded CSI,'' \emph{IEEE Trans. Wireless Commun.}, vol. 20, no. 4, pp. 2487-2501, Apr. 2021.

\bibitem{cascaded}
G. Zhou et al., ``A framework of robust transmission design for IRS-aided MISO communications with imperfect cascaded channels,''  \emph{IEEE Trans. Signal Process.}, vol. 68, pp. 5092-5106, 2020.



\bibitem{MAMag}
L. Zhu, W. Ma, and R. Zhang, “Movable antennas for wireless communication: Opportunities and challenges,” \emph{IEEE Commun. Mag.}, vol. 62, no. 6, pp. 114-120, Jun. 2024.




\bibitem{MATWC}
W. Ma, L. Zhu, and R. Zhang, ``MIMO capacity characterization for movable antenna systems,'' \emph{IEEE Trans. Wireless Commun.}, vol. 23, no. 4, pp. 3392-3407, Apr. 2024.


\bibitem{FAS}
K. -K. Wong et al., ``Fluid antenna systems,'' \emph{IEEE Trans. Wireless Commun.}, vol. 20, no. 3, pp. 1950-1962, March 2021.



\bibitem{Lipeng}
L. Zhu, W. Ma, and R. Zhang, ``Modeling and performance analysis for movable antenna enabled wireless communications,'' \emph{IEEE Trans. Wireless Commun.}, vol. 23, no. 6, pp. 6234-6250, Jun. 2024.



\bibitem{Weidong}
W. Mei, X. Wei, B. Ning, Z. Chen, and R. Zhang, ``Movable-antenna position optimization: A graph-based approach,'' \emph{IEEE Wireless Commun. Lett.}, vol. 13, no. 7, pp. 1853-1857, Jul. 2024.


\bibitem{QQWu}
H. Wang, Q. Wu, and W. Chen,  ``Movable antenna enabled interference network: Joint antenna position and beamforming design,'' \emph{EEE Wireless Commun. Lett.}, vol. 13, no. 9, pp. 2517–2521, Sep. 2024.



\bibitem{MAthree}
Y. Zhou, W. Chen, Q. Wu, X. Zhu, and N. Cheng, ``Movable antenna empowered downlink NOMA systems: Power allocation and antenna position optimization,'' \emph{IEEE Wireless Commun. Lett.}, vol. 13, no. 10, pp. 2772-2776, Oct. 2024.


\bibitem{MA-TMC}
G. Hu et al.,  ``Movable antennas-assisted secure transmission without eavesdroppers’ instantaneous CSI,'' \emph{IEEE Trans. Mobile Comput.}, vol. 23, no. 12, pp. 14263-14279, Dec. 2024.


\bibitem{MoveDelay}
H. Wang et al., ``Throughput maximization for movable antenna systems with movement delay consideration," [Online]. Available: https://arxiv.org/abs/2411.13785, 2024.




\bibitem{PSOMA}
Z. Xiao, X. Pi, L. Zhu, X.-G. Xia, and R. Zhang, ``Multiuser communications with movable-antenna base station: Joint antenna positioning, receive combining, and power control,'' \emph{IEEE Trans. Wireless Commun.}, vol. 23, no. 12, pp. 19744-19759, Dec. 2024. 

\bibitem{TwoScale}
G. Hu et al., ``Two-timescale design for movable antenna array-enabled multiuser uplink communications,'' \emph{IEEE Trans. Veh. Technol.}, doi: 10.1109/TVT.2024.3485647, 2024.



\bibitem{Pengcheng}
P. Chen et al., ``Movable antenna-enhanced wireless powered mobile edge computing systems,'' \emph{IEEE Internet Things J.}, vol. 11, no. 21, pp. 35505-35518, Nov. 2024.




\bibitem{ZHOUMASR}
C. Zhou, B. Lyu, C. You, and Z. Liu, ``Movable antenna enabled symbiotic radio systems: An opportunity for mutualism,'' \emph{IEEE Wireless Commun. Lett.}, vol. 13, no. 10, pp. 2752-2756, Oct. 2024.

\bibitem{guanjiayu}
 J. Guan et al., ``Secure transmission for movable antennas empowered cell-free symbiotic radio communications,'' in \emph{Proc. the 16th Int. Conf. Wireless Commun. Signal Process. },  Hefei, China, Oct. 2024, pp. 1-7.


\bibitem{Bernstein-Type}
 K. Wang, A. M. So, T. Chang, W. Ma, and C. Chi, ``Outage constrained
robust transmit optimization for multiuser miso downlinks: Tractable
approximations by conic optimization,'' \emph{IEEE Trans. Signal Process.}, vol. 62, no. 21, pp. 5690–5705, Nov. 2014.

\bibitem{Uncertainty}
 X. Yu, D. Xu, Y. Sun, D. W. K. Ng, and R. Schober, ``Robust and secure wireless communications via intelligent reflecting surfaces,'' \emph{IEEE J. Sel. Areas Commun.}, vol. 38, no. 11, pp. 2637-2652, Nov. 2020.


\bibitem{HPSO}
Z. Yu, Z. Si, X. Li, D. Wang, and H. Song, ``A novel hybrid particle swarm optimization algorithm for path planning of UAVs,'' \emph{IEEE Internet Things J.}, vol. 9, no. 22, pp. 22547-22558, Nov. 2022.

\bibitem{ssp}
S. Boyd et al., \emph{Linear Matrix Inequalities in System and Control Theory.} Philadelphia, PA, USA: SIAM, 1994.

\bibitem{lemmatwo}
E. A. Gharavol et al., ``The sign-definiteness lemma and
its applications to robust transceiver optimization for multiuser MIMO systems,'' \emph{IEEE Trans. Signal Process.}, vol. 61, no. 2, pp. 238–252, Jan. 2013.

\bibitem{logzeng}
Y. Tang, J. Xiong, D. Ma, and X. Zhang, ``Robust artificial noise aided transmit design for MISO wiretap channels with channel uncertainty,''
\emph{IEEE Commun. Lett.}, vol. 17, no. 11, pp. 2096–2099, Nov. 2013.

\bibitem{CVX}
S. Boyd et al., \emph{Convex Optimization}. 	Cambridge University Press, 2004.


\bibitem{Two-timescale}
C. Hu, L. Dai, S. Han, and X. Wang, ``Two-timescale channel estimation for reconfigurable intelligent surface aided wireless communications,'' \emph{IEEE Trans. Commun.}, vol. 69, no. 11, pp. 7736-7747, Nov. 2021.



\end{thebibliography}
\end{document}